\documentclass[english]{IEEEtran}
\usepackage[T1]{fontenc}
\usepackage[latin9]{inputenc}
\usepackage{xcolor}
\usepackage{pdfcolmk}
\usepackage{multirow}
\usepackage{amsthm}
\usepackage{amsmath}
\usepackage{amssymb}
\usepackage{graphicx}
\usepackage{esint}
\PassOptionsToPackage{normalem}{ulem}
\usepackage{ulem}

\makeatletter

\providecommand{\tabularnewline}{\\}
\providecolor{lyxadded}{rgb}{0,0,1}
\providecolor{lyxdeleted}{rgb}{1,0,0}

  \theoremstyle{plain}
  \newtheorem{lem}{\protect\lemmaname}
  \theoremstyle{remark}
  \newtheorem{rem}{\protect\remarkname}
  \theoremstyle{definition}
  \newtheorem{defn}{\protect\definitionname}
  \theoremstyle{plain}
  \newtheorem{thm}{\protect\theoremname}
  \theoremstyle{plain}
  \newtheorem{cor}{\protect\corollaryname}

\usepackage{cite}
\usepackage{times}

\newtheorem{assumption}{Assumption}

\newtheorem{condition}{Condition}
\usepackage{subfig}

\makeatother

\usepackage{babel}
\providecommand{\corollaryname}{Corollary}
\providecommand{\definitionname}{Definition}
\providecommand{\lemmaname}{Lemma}
\providecommand{\remarkname}{Remark}
\providecommand{\theoremname}{Theorem}

\begin{document}

\title{Asymaptotic Scaling Laws of Wireless Adhoc Network with Physical
Layer Caching}

\author{{\normalsize{}An Liu, }\textit{\normalsize{}Member IEEE}{\normalsize{},
and Vincent Lau,}\textit{\normalsize{} Fellow IEEE}{\normalsize{},\\Department
of Electronic and Computer Engineering, Hong Kong University of Science
and Technology}}
\maketitle
\begin{abstract}
We propose a \textit{physical layer (PHY) caching} scheme for wireless
adhoc networks. The PHY caching exploits \textit{cache-assisted multihop
gain} and \textit{cache-induced dual-layer CoMP gain}, which substantially
improves the throughput of wireless adhoc networks. In particular,
the PHY caching scheme contains a novel PHY transmission mode called
the \textit{cache-induced dual-layer CoMP} which can support homogeneous
opportunistic CoMP in the wireless adhoc network. Compared with traditional
per-node throughput scaling results of $\Theta\left(1/\sqrt{N}\right)$,
we can achieve $O(1)$ per node throughput for a cached wireless adhoc
network with $N$ nodes. Moreover, we analyze the throughput of the
PHY caching scheme for regular wireless adhoc networks and study the
impact of various system parameters on the PHY caching gain.\end{abstract}


\section{Introduction}

Recently, wireless caching has been proposed as a cost-effective solution
to handle the high traffic rate caused by content delivery applications
\cite{Goebbels_IJCS10_Smartcaching,Caire_INFOCOM12_femtocache}. By
exploiting the fact that content is \textquotedblleft cachable\textquotedblright ,
wireless nodes can cache some popular content during off-peak hours
(\textit{cache initialization phase}), in order to reduce traffic
rate at peak hours (\textit{content delivery phase}). Caching has
been widely used in wired networks such as fixed line P2P systems
\cite{Kozat_TOM09_P2P} and content distribution networks (CDN) \cite{Shen_TOM04_CDN}.
One key difference between wireless and wired networks is that the
performance of wireless networks is fundamentally limited by the interference.
However, this unique feature of wireless networks is not fully exploited
in existing wireless caching schemes. Recently, more advanced interference
mitigation techniques such as coordinated multipoint (CoMP) transmission
\cite{Xiaohu_TWC11_COMIMO} have been proposed. Conventionally, the
CoMP technique requires high capacity backhaul for payload exchange
between the transmitting nodes. However, backhaul connectivity to
the transmitting node may not be available in situations such as wireless
adhoc networks. An interesting question is that, can we exploit wireless
caching to achieve CoMP gain in wireless adhoc networks without payload
backhaul connections between the nodes? 

\begin{figure}
\begin{centering}
\textsf{\includegraphics[clip,width=88mm]{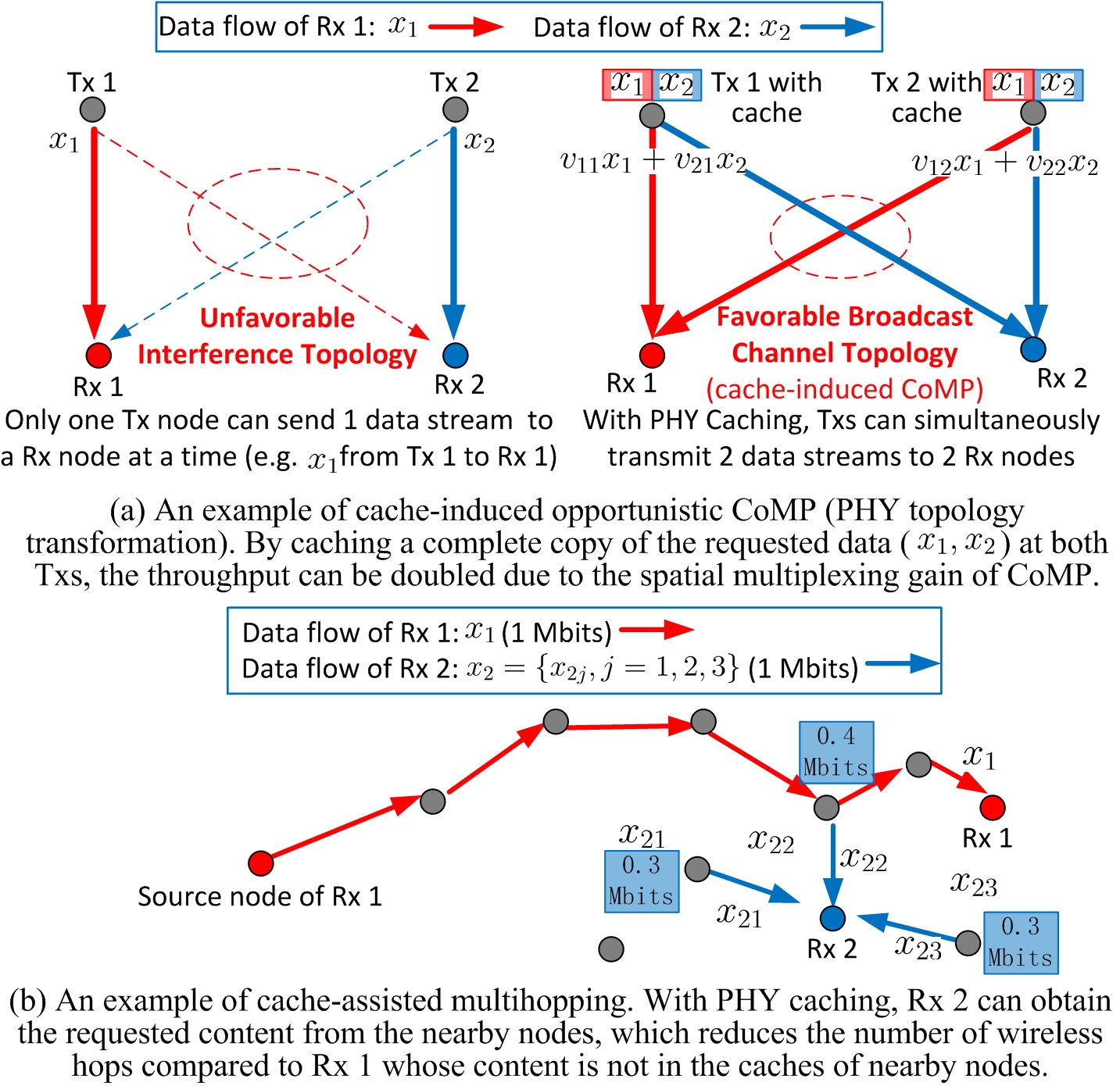}}
\par\end{centering}

\protect\caption{\label{fig:TopoChange}{\small{}An illustration of cache-induced opportunistic
CoMP and cache-assisted multihopping. In Fig. \ref{fig:TopoChange}-(a),
$\mathbf{v}_{1}=\left[v{}_{11},v_{12}\right]^{T}$ and $\mathbf{v}_{2}=\left[v{}_{21},v_{22}\right]^{T}$
are the beamforming vectors for Rx 1 and Rx 2, respectively.}}
\end{figure}

In this paper, we propose a \textit{PHY caching} scheme to achieve
both \textit{cache-induced opportunistic CoMP} and \textit{cache-assisted
multihopping} in wireless adhoc network as elaborated below. 
\begin{itemize}
\item \textbf{Cached-induced opportunistic CoMP:} If the content accessed
by several nodes exists simultaneously at the nearby nodes (i.e.,
each nearby node has a complete copy of the requested content), the
nearby nodes can engage in CoMP and enjoy CoMP gain. In this way,
we can opportunistically transform the \textit{interference network
topology} into a more favorable \textit{MIMO broadcast channel topology},
as illustrated in Fig. \ref{fig:TopoChange}-(a), and this is referred
to as cache-induced opportunistic CoMP. 
\item \textbf{Cached-assisted multihopping:} If the content requested by
a node is distributed{\small{} }in the caches of several nearby nodes
(i.e., each nearby node has a different portion of the requested content),
this node can directly obtain the requested content from the nearby
nodes, which will significantly reduce the number of hops from the
source nodes to the destination node, as illustrated in Fig. \ref{fig:TopoChange}-(b).
This is referred to as cache-assisted multihopping.
\end{itemize}

We are interested in studying the benefits of PHY caching and its
impact on the throughput scaling law of \textit{cache-assisted wireless
adhoc networks}. Some related works are reviewed below. \cite{Niesen_TIT12_FLcaching}
proposed coded caching schemes that can create coded multicast opportunities.
A proactive caching paradigm was proposed in \cite{Debbah_CoM14_proactivecache}
to exploit both the spatial and social structure of the wireless networks.
\cite{Gitzenis_TIT13_wirelesscache} studied the joint optimization
of cache content replication and routing in a \textit{regular network}
and identified the throughput scaling laws for various regimes. Recently,
a number of works have studied the fundamental tradeoff or scaling
laws in wireless device-to-device (D2D) caching networks \cite{Caire_arxiv13_D2Dcaching,Caire_arxiv13_d2dcachingtradeoff,Caire_arxiv13_d2dcachingtutorial,Altieri_arxiv14_d2dcaching,Jeon_ICC15_D2Dcaching}.
For example, \cite{Caire_arxiv13_D2Dcaching} investigated the fundamental
limits of distributed caching in D2D wireless networks and the combined
effect of coded multicast gain and spatial reuse. \cite{Caire_arxiv13_d2dcachingtradeoff}
characterized the optimal throughput-outage tradeoff in D2D caching
networks in terms of tight scaling laws for various regimes. \cite{Caire_arxiv13_d2dcachingtutorial}
presented a tutorial for the schemes and the recent results on the
throughput scaling laws of D2D caching networks. A stochastic geometry
framework, which incorporates the effects of random distributed users,
user clustering, fading channel and interference, was proposed in
\cite{Altieri_arxiv14_d2dcaching} to study the tradeoff between the
fraction of video requests served through D2D and the average rate.
Note that the scaling laws studied in the above works and in this
paper are different in many aspects. First, the existing caching schemes
do not consider cache-induced CoMP among the nodes. In this paper,
the roles of PHY caching include both the \textit{cache-assisted multihopping}
and the \textit{cache-induced dual-layer} \textit{CoMP}. Second, many
theoretical results on scaling laws in wireless caching networks are
based on the simple \textquotedblleft protocol model\textquotedblright{}
\cite{GuptaKumar,Caire_arxiv13_d2dcachingtradeoff} without considering
the effect of PHY channel. Although some existing works such as \cite{Altieri_arxiv14_d2dcaching}
considered PHY channel model, none of them studied the coupling between
caching and PHY transmission. In our analysis, we consider both the
effect of PHY channel model and the coupling between caching and PHY
transmission (e.g., the PHY transmission modes in our design depend
on the cache mode of the requested file). Note that the concept of
cached-induced opportunistic CoMP was first introduced in our previous
works \cite{Liu_TSP14_CacheRelay,Liu_TSP13_CacheIFN} for cellular
networks. However, the cached-induced opportunistic CoMP schemes in
\cite{Liu_TSP14_CacheRelay,Liu_TSP13_CacheIFN} cannot be directly
applied to wireless adhoc networks where a node may serve both as
a transmitter that provides content and a receiver that requests content.
As a result, we propose a new cache-induced \textit{dual-layer} CoMP
to support cached-induced opportunistic CoMP in wireless adhoc networks.
To the best of our knowledge, a complete understanding of the role
of caching in wireless networks is still missing, especially for the
inter-play between the cache-assisted multihop gain and the cache-induced
CoMP gains. The main contributions of this paper are as follows.
\begin{itemize}
\item \textbf{PHY Caching for Multihopping and CoMP Gains: }We propose a
PHY caching design to exploit both\textit{ cache-assisted multihopping
}and\textit{ cache-induced dual-layer CoMP} in wireless adhoc networks. 
\item \textbf{Closed-form Performance Analysis for a Regular Wireless Adhoc
Network: }We derive closed-form expression of the per node throughput
for the regular wireless adhoc network defined in Section \ref{sec:Performance-Analysis}.
As a comparison, we also analyze the performance of a baseline\textit{
multihop caching scheme} which only achieves cache-assisted multihop
gain. We quantify the cache-induced dual-layer CoMP gain (i.e., the
throughput gain of the proposed PHY caching over the multihop caching)
and study the impact of various system parameters on the performance.
\item \textbf{Throughput Scaling of Cache-assisted Wireless Adhoc Network:
}Under Zipf popularity distribution \cite{Breslau_INFOCOM99_ZipfLaw,Yamakami_PDCAT06_Zipflaw},
we show that\textbf{ }the per node throughput of the cache-assisted
wireless adhoc network with $N$ nodes scales from $\Theta\left(1/\sqrt{N}\right)$
to $\Theta\left(1\right)$ depending on the Zipf parameter $\tau$
and cache size. 
\end{itemize}

The rest of the paper is organized as follows. In Section \ref{sec:System-Model},
we introduce the system model. The PHY caching scheme is elaborated
in Section \ref{sec:Order-wise-Optimal-Control}. The performance
analysis for the regular wireless adhoc network is given in Section
\ref{sec:Performance-Analysis}. The throughput scaling law of the
proposed PHY caching is derived in Section \ref{sec:Scaling-Laws-of}
for general cache-assisted wireless adhoc network. Some numerical
results are presented in Section \ref{sec:Numerical-Results-and}.
The conclusion is given in \ref{sec:Conclusion}. The key notations
are summarized in Table \ref{tab:notations}.

\begin{table}
\begin{centering}
\begin{tabular}{|l|c|}
\hline 
{\footnotesize{}$l_{n}$} & {\footnotesize{}Index of the file requested by node $n$}\tabularnewline
\hline 
{\footnotesize{}$q_{l}$} & {\footnotesize{}Cache content replication variable for file $l$}\tabularnewline
\hline 
{\footnotesize{}$p_{l}$} & {\footnotesize{}Probability of requesting file $l$}\tabularnewline
\hline 
{\footnotesize{}$\mathcal{N}_{C}^{1}$ ($\mathcal{N}_{C}^{2}$) } & {\footnotesize{}The set of all layer 1 (2) CoMP Tx nodes}\tabularnewline
\hline 
{\footnotesize{}$\Omega\left(\mathbf{q}\right)$ ($\overline{\Omega}\left(\mathbf{q}\right)$)} & {\footnotesize{}The set of files with multihop (CoMP) cache mode}\tabularnewline
\hline 
{\footnotesize{}$W_{b}$ ($W_{c}$)} & {\footnotesize{}Bandwidth of multihop (each CoMP) band}\tabularnewline
\hline 
{\footnotesize{}$\mathcal{B}_{n}$} & {\footnotesize{}Source node set of node n}\tabularnewline
\hline 
{\footnotesize{}$\mathcal{G}_{1,j}^{\textrm{Tx}}$ ($\mathcal{G}_{1,j}^{\textrm{Rx}}$)} & {\footnotesize{}$j$-th layer 1 CoMP transmit (receive) cluster}\tabularnewline
\hline 
{\footnotesize{}$N_{c}$} & {\footnotesize{}CoMP cluster size}\tabularnewline
\hline 
\multirow{1}{*}{{\footnotesize{}$\Gamma_{B}\left(\mathbf{q}\right)$ ($\Gamma_{A}\left(\mathbf{q}\right)$)}} & {\footnotesize{}Per node throughput under multihop (PHY) caching }\tabularnewline
\hline 
\end{tabular}
\par\end{centering}

\protect\caption{\label{tab:notations} {\small{}Key notations}}
\end{table}

\section{Cache-assisted Wireless Adhoc Network Model\label{sec:System-Model}}

\begin{figure}
\begin{centering}
\textsf{\includegraphics[clip,width=75mm]{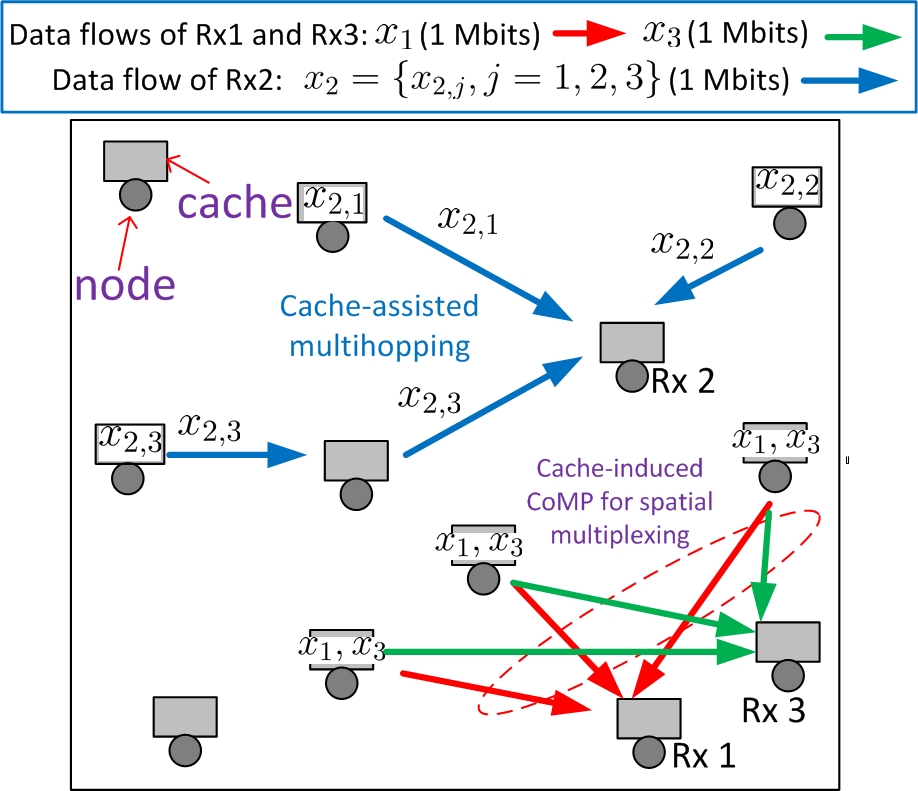}}
\par\end{centering}

\protect\caption{\label{fig:system_model}{\small{}Architecture of the cache-assisted
wireless adhoc network.}}
\end{figure}

Consider a cache-assisted wireless adhoc network with $N$ nodes randomly
placed on a square of area $Nr_{0}^{2}$ as illustrated in Fig. \ref{fig:system_model}.
Each node has an average transmit power budget of $P$ and a cache
of size $B_{C}$ bits. We have the following assumption on the node
placement.

\begin{assumption}[Node Placement]\label{asm:Place}The node placement
satisfies the following conditions.
\begin{enumerate}
\item The distance between any two nodes is no less than some constant $r_{\textrm{min}}>0$.
\item For any point in the square occupied by the network, the distance
between this point and nearest node is no more than some constant
$r_{\textrm{max}}>0$. 
\end{enumerate}
\end{assumption}

Assumption \ref{asm:Place}-1) is always satisfied in practice. Assumption
\ref{asm:Place}-2) is to avoid the case when some nodes concentrate
in a few isolated spots. Such case is undesired because a node in
an isolated spot cannot be served whenever it requests content which
is not in the caches of the nodes in the same spot.

In cache-assisted wireless adhoc network, the nodes request data (e.g.,
music or video) from a set of content files indexed by $l\in\mathcal{L}=\left\{ 1,2,...,L\right\} $.
The size of the $l$-th file is $F_{l}$ bits and we assume $F_{l}=\Theta\left(B_{C}\right),\forall l$.
There are two phases during the operation of cache-assisted wireless
adhoc network, namely the\textit{ cache initialization phase} and
the \textit{content delivery phase}. 

In the cache initialization phase, each node caches a portion of $q_{l}F_{l}$
(possibly encoded) bits of the $l$-th content file ($\forall l$)
, where $\mathbf{q}=\left[q_{1},...,q_{L}\right]^{T}$ (with $q_{l}\in\left[0,1\right]$
and $\sum_{l=1}^{L}q_{l}F_{l}\leq B_{C}$) are called \textit{cache
content replication vector}. The specific cache data structure at
each node and the algorithm to find $\mathbf{q}$ will be elaborated
in Section \ref{sub:MDS-Cache-Encoding} and Section \ref{sec:Scaling-Laws-of},
respectively. Since the popularity of content files change very slowly
(e.g., new movies are usually posted on a weekly or monthly timescale),
the cache update overhead in the cache initialization phase is usually
small \cite{Gitzenis_TIT13_wirelesscache}. This is a reasonable assumption
widely used in the literature.\textbf{ }However, the impact of \textquotedblleft cache
initialization phase\textquotedblright{} is worthy of further investigation,
especially for the case when the popularity of content files changes
more frequently. 

In the content delivery phase, if the content requested by node $n$
is not in its own cache, it will obtain the requested content from
a subset of other nodes that cache the requested content via cache-induced
CoMP or cache-assisted multihopping. For example, in Fig. \ref{fig:system_model},
the content accessed by Rx 1 and Rx 3 exists simultaneously at the
cache of the nearby nodes and thus they can be served by the nearby
nodes using cache-induced CoMP, enjoying spatial multiplexing gains.
The content requested by Rx 2 is distributed in the caches of the
nearby nodes and thus it is served by the nearby nodes using cache-assisted
multihopping. For convenience, let $l_{n}$ denote the index of the
file requested by node $n$ and let $\vec{l}=\left\{ l_{1},...,l_{N}\right\} $
denote the \textit{user request profile} (URP). The URP process $\vec{l}\left(t\right)$
is an ergodic random process. Specifically, each node independently
accesses the $l$-th content file with probability $p_{l}$, where
probability mass function $\mathbf{p}=\left[p_{1},...,p_{L}\right]$
represents the popularity of the content files.

We assume $B_{C}<\sum_{l=1}^{L}F_{l}$ to avoid the trivial case when
each node can cache all content files. Furthermore, we assume $NB_{C}>\sum_{l=1}^{L}F_{l}$
so that there is at least one complete copy of each content file in
the caches of the entire network. We use similar channel model as
in \cite{Tse_IT07_CapscalingHMIMO,Niesen_TIT10_CSadhoc}. 

\begin{assumption}[Channel Model]\label{asm:channel}The wireless
link between two nodes is modeled by a flat block fading channel with
bandwidth $W$. The channel coefficient between node $n$ and $n^{'}$
at time slot $t$ is
\[
h_{n^{'},n}\left(t\right)=\left(r_{n^{'},n}\right)^{-\alpha/2}\exp\left(j\theta_{n^{'},n}\left(t\right)\right),
\]
where $r_{n^{'},n}$ is the distance between node $n$ and $n^{'}$,
$\theta_{n^{'},n}\left(t\right)$ is the random phase at time $t$,
and $\alpha>2$ is the path loss exponent. Moreover, $\theta_{n^{'},n}\left(t\right)$
are i.i.d. (w.r.t. the node index $n^{'},n$ and time index $t$)
with uniform distribution on $\left[0,2\pi\right]$. 

\end{assumption}

At each node, the received signal is also corrupted by a circularly
symmetric Gaussian noise. Without loss of generality, the spectral
density of the noise is normalized to be 1.

\section{PHY Caching for Wireless Adhoc Networks\label{sec:Order-wise-Optimal-Control}}

We first outline the key components of the PHY caching scheme. Then
we elaborate each component.

\subsection{Key Components of the PHY Caching Scheme}

\begin{figure}
\begin{centering}
\textsf{\includegraphics[clip,width=80mm]{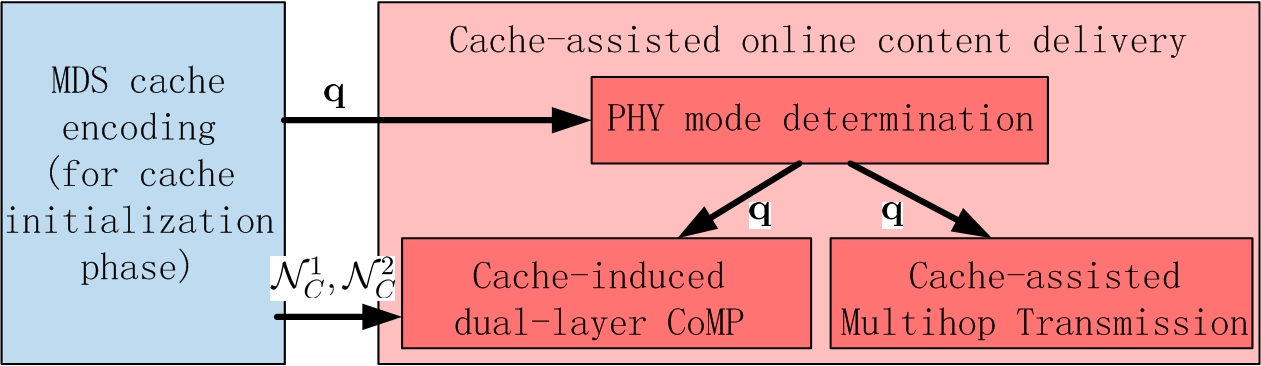}}
\par\end{centering}

\protect\caption{\label{fig:PHYcache_comm}{\small{}Components of PHY caching and their
inter-relationship.}}
\end{figure}

The components of the proposed PHY caching scheme and their inter-relationship
are illustrated in Fig. \ref{fig:PHYcache_comm}. There are two major
components: the\textit{ maximum distance separable} \textit{(MDS)
cache encoding} working in the cache initiation phase and the \textit{cache-assisted
online content delivery} working in the content delivery phase. The
MDS cache encoding component converts the content files into coded
segments and decides how to cache the coded segments at each node.
The cache-assisted online content delivery component exploits the
coded segments cached at each node to achieve both cache-induced CoMP
gain and cache-assisted multihop gain. Specifically, there are two
PHY transmission modes, namely, \textit{cache-assisted multihop transmission}
and c\textit{ache-induced dual-layer CoMP transmission}. At each node,
a PHY mode determination component first determines the transmission
mode. Then each node obtains the requested file using the corresponding
transmission mode.

\subsection{Offline MDS Cache Encoding for Cache Initialization Phase\label{sub:MDS-Cache-Encoding}}

\begin{figure}
\begin{centering}
\textsf{\includegraphics[clip,width=85mm]{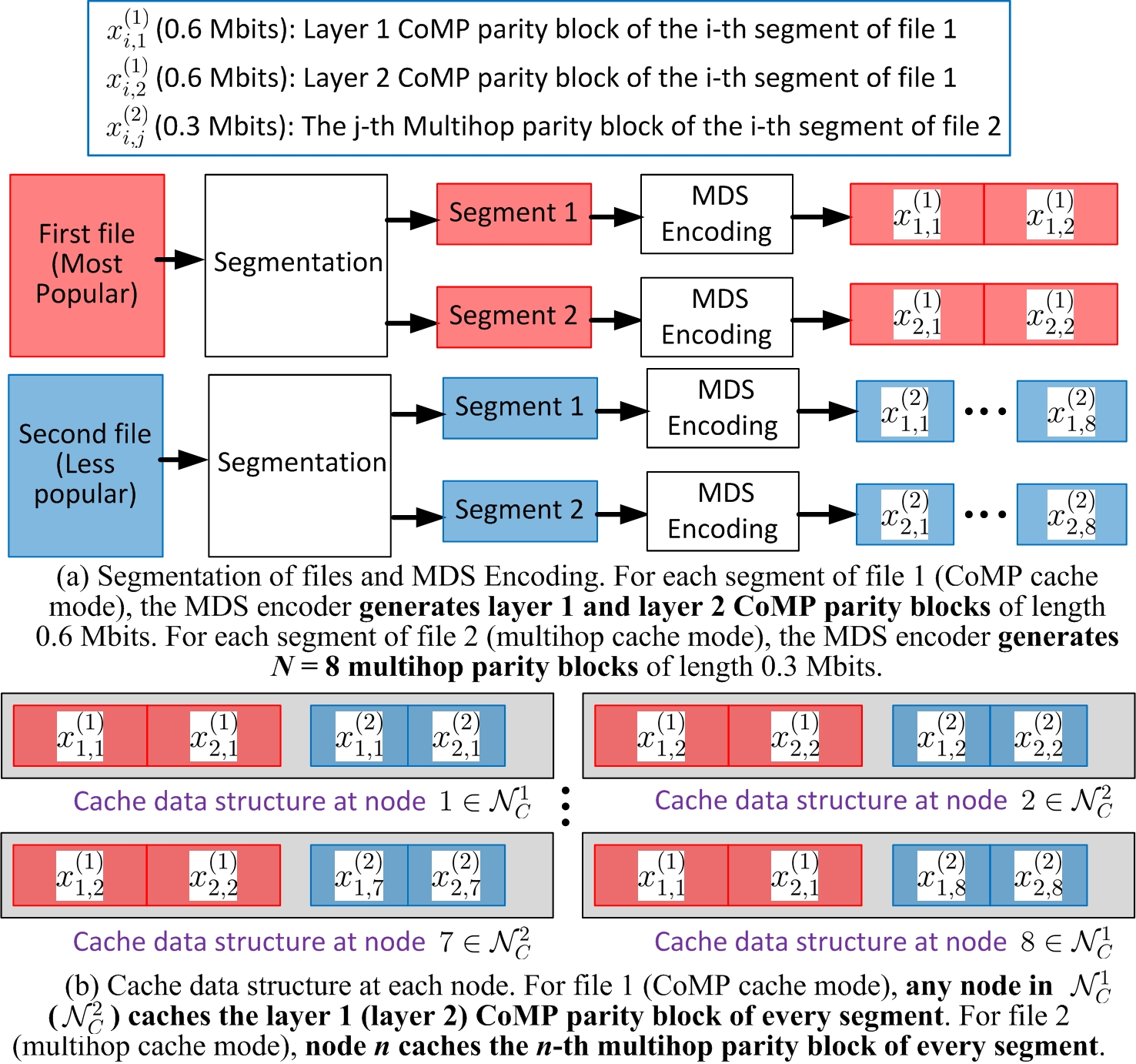}}
\par\end{centering}

\protect\caption{\label{fig:MDS_cache_struc}{\small{}An illustration of MDS cache
encoding. The network has $N=8$ nodes as shown in Fig. \ref{fig:cachePHYmodes}.
There are $2$ content files with $q_{1}=0.6$ and $q_{2}=0.3$. The
size of each file is $2$ Mbits and the segment size is $L_{S}=1$
Mbits. The cache size $B_{C}$ is 1.8 Mbits.}}
\end{figure}

The cache-assisted multihopping and cache-induced dual-layer CoMP
have conflicting requirements on the caching scheme. For the former,
it is better to cache different content at different nodes so that
the number of hops from the source to destination nodes can be minimized.
For the later, the cache of the nearby transmitting nodes should store
the same contents to support \textit{spatial multiplexing}. Here,
spatial multiplexing refers to simultaneous transmission of multiple
data streams from a set of nearby serving nodes (which have the requested
content in their caches) to a set of nearby requesting nodes using
CoMP. We propose an \textit{MDS cache encoding scheme} to strike a
balance between these two conflicting goals. \vspace{0.1in}

\textbf{\textit{MDS Cache Encoding Scheme}} (parameterized by a \textit{cache
content replication vector} $\mathbf{q}=\left[q_{1},...,q_{L}\right]^{T}$)

{\small{}S}\textbf{\small{}tep 1 (Nodes Partitioning): }{\small{}Partition
the nodes into two non-overlapping subsets $\mathcal{N}_{C}^{1}$
and $\mathcal{N}_{C}^{2}$ using the following algorithm. First, node
1 marks itself as a node in $\mathcal{N}_{C}^{1}$ and broadcasts
a MARK message containing a }\textit{\small{}mark bit}{\small{} to
the adjacent nodes. The mark bit is initialized to be bit 0. When
a node receives a MARK message with SINR larger than a threshold $\gamma_{0}$
for the first time, if the mark bit is 0 (1), it marks itself as a
node in $\mathcal{N}_{C}^{2}$ ($\mathcal{N}_{C}^{1}$ ) and sends
a MARK message with mark bit 1 (0) to the adjacent nodes. After each
node receives a MARK message (with SINR larger than $\gamma_{0}$)
for at least once, the nodes have been partitioned into two non-overlapping
subsets $\mathcal{N}_{C}^{1}$ and $\mathcal{N}_{C}^{2}$. The threshold
$\gamma_{0}$ can be chosen according to the rules used in the existing
neighbor discovery algorithms for wireless adhoc networks.}{\small \par}

{\small{}S}\textbf{\small{}tep 2 (MDS Encoding and Cache Modes Determination):}{\small{}
Each file is divided into segments of $L_{S}$ bits. Each segment
is encoded using a MDS rateless code as shown in Fig. \ref{fig:MDS_cache_struc}-(a).
If $q_{l}\geq0.5$, the cache mode for the $l$-th content file is
set to be }\textit{\small{}CoMP cache mode}{\small{}. In this case,
for each segment of the $l$-th file, the MDS encoder first generates
$2q_{l}L_{S}$ parity bits from the $L_{S}$ information bits of the
original segment. Then the first $q_{l}L_{S}$ parity bits form the
}\textit{\small{}layer 1 CoMP parity block}{\small{} and the last
$q_{l}L_{S}$ parity bits form the }\textit{\small{}layer 2 CoMP parity
block}{\small{} of this segment, as illustrated in Fig. \ref{fig:MDS_cache_struc}-(a)
for the first file. If $q_{l}<0.5$, the cache mode for the $l$-th
content file is set to be }\textit{\small{}multihop cache mode }{\small{}and
the MDS encoder generates $N$ }\textit{\small{}multihop parity blocks}{\small{}
of length $q_{l}L_{S}$ for each segment of the $l$-th file as illustrated
in Fig. \ref{fig:MDS_cache_struc}-(a) for the second file. Define
$\Omega\left(\mathbf{q}\right)\triangleq\left\{ l:\: q_{l}<0.5\right\} $
as the set of files associated with multihop cache mode and $\overline{\Omega}\left(\mathbf{q}\right)\triangleq\left\{ l:\: q_{l}\geq0.5\right\} $
as the set of files associated with CoMP cache mode. }{\small \par}

\textbf{\small{}Step 3 (Offline Cache Initialization):}{\small{} For
$l=1,..,L$, if $l\in\Omega\left(\mathbf{q}\right)$, then the cache
of node $n$ is initialized with the $n$-th multihop parity block
for each segment of the $l$-th file. If $l\in\overline{\Omega}\left(\mathbf{q}\right)$,
then the caches of the nodes in $\mathcal{N}_{C}^{1}$ and $\mathcal{N}_{C}^{2}$
are initialized with the layer 1 and layer 2 CoMP parity blocks respectively.
}\vspace{0.1in}

\begin{figure}
\begin{centering}
\textsf{\includegraphics[clip,width=85mm]{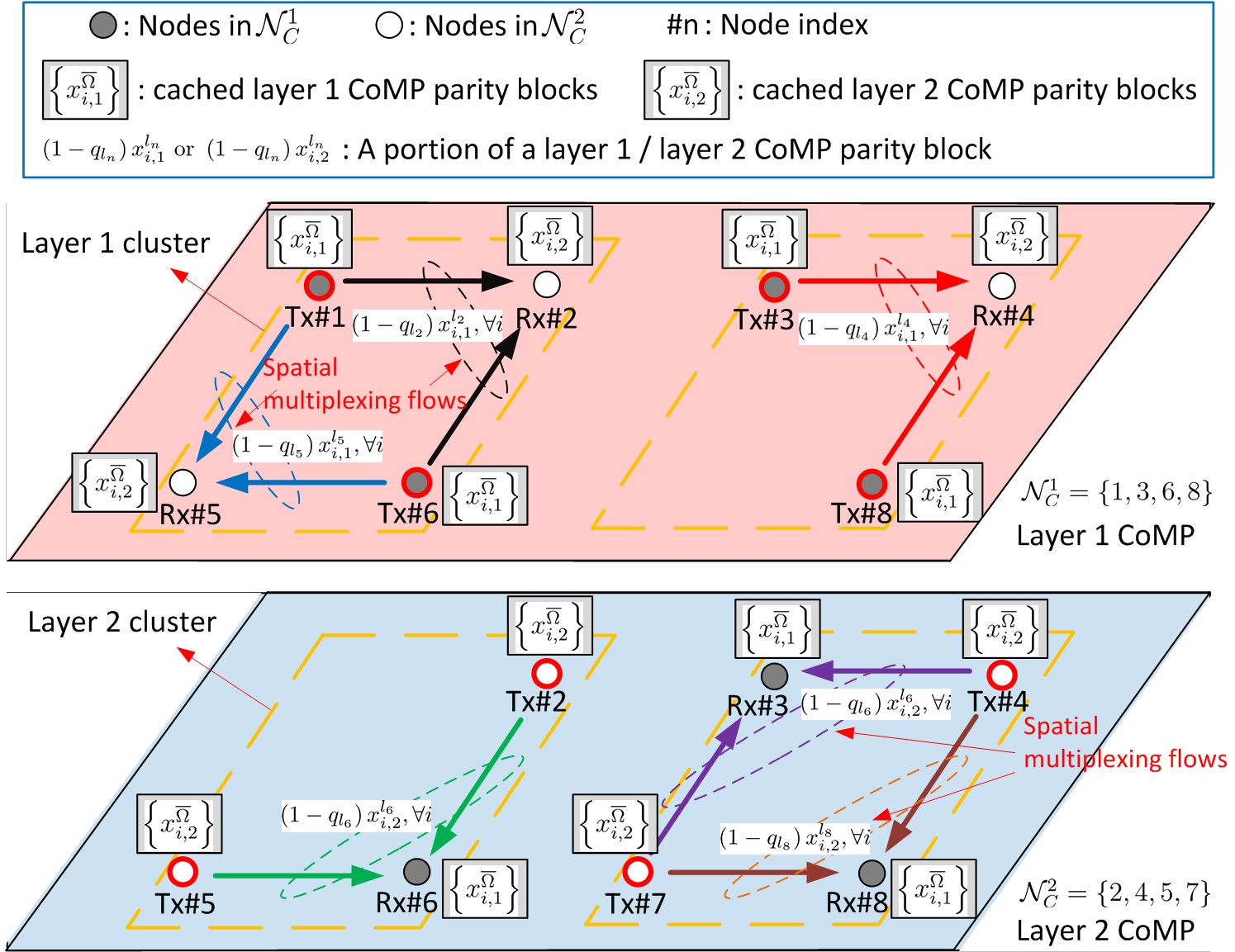}}
\par\end{centering}

\protect\caption{\label{fig:cachePHYmodes}{\small{}An illustration of cache-induced
dual-layer CoMP, where each node in $\left\{ 2,3,4,5,6,8\right\} $
requests a file associated with CoMP cache mode.}}
\end{figure}

The MDS cache encoding scheme has several benefits. First, the parity
blocks of the requested segment can be received in any order without
protocol overheads of reassembly due to the property of MDS codes.
Second, when a node requests a file associated with multihop cache
mode, it can always obtain the parity blocks required to decode this
file from the nearest $\left\lceil 1/q_{l}\right\rceil -1$ nodes.
Third, the introduction of CoMP cache mode for popular files facilitates
the design of cache-induced dual-layer CoMP. To support homogeneous
opportunistic CoMP in the wireless adhoc network, we propose a \textit{cache-induced
dual-layer CoMP} as illustrated in Fig. \ref{fig:cachePHYmodes}.
Specifically, the nodes in the adhoc network are partitioned into
two non-overlapping subsets $\mathcal{N}_{C}^{1}$ and $\mathcal{N}_{C}^{2}$
such that $\mathcal{N}_{C}^{1}\cap\mathcal{N}_{C}^{2}=\emptyset$
as illustrated in Step 1. The nodes in $\mathcal{N}_{C}^{1}$ ($\mathcal{N}_{C}^{2}$)
cache the layer 1 (layer 2) CoMP parity blocks so that requesting
nodes in $\mathcal{N}_{C}^{2}$ can be served with layer 1 CoMP transmission
from nodes in $\mathcal{N}_{C}^{1}$, enjoying spatial multiplexing
gains and vice versa for layer 2. For example, in Fig. \ref{fig:cachePHYmodes},
the nodes in $\mathcal{N}_{C}^{1}=\left\{ 1,3,6,8\right\} $ ($\mathcal{N}_{C}^{2}=\left\{ 2,4,5,7\right\} $)
cache $\left\{ x_{i,1}^{\overline{\Omega}}\right\} $ ($\left\{ x_{i,2}^{\overline{\Omega}}\right\} $),
which denotes the set of all layer 1 (2) parity blocks of all files
$l\in\overline{\Omega}\left(\mathbf{q}\right)$ associated with the
CoMP cache mode. The requesting nodes $2,4,5$ belong to $\mathcal{N}_{C}^{2}$
and thus they are served by the nodes in $\mathcal{N}_{C}^{1}=\left\{ 1,3,6,8\right\} $
using the layer 1 CoMP as illustrated in the red plane. The requesting
nodes $3,6,8$ belong to $\mathcal{N}_{C}^{1}$ and thus they are
served by the nodes in $\mathcal{N}_{C}^{2}=\left\{ 2,4,5,7\right\} $
using the layer 2 CoMP as illustrated in the blue plane. 

The cache-assisted multihop gain and the cache-induced dual-layer
CoMP gain depends heavily on the choice of the cache content replication
vector $\mathbf{q}$. In Theorem \ref{thm:Order-optimal-pR}, we will
give an \textit{order-optimal} cache content replication vector $\mathbf{q}$
to maximize the order of the per node throughput. The order-optimal
$\mathbf{q}$ is calculated offline based on the content popularity
$\mathbf{p}$.

\subsection{Frequency Planning for Interference Mitigation}

The interference in adhoc networks is mitigated using the frequency
planning technique. The bandwidth $W$ is divided into three bands,
namely, the \textit{multihop band} with size $W_{b}$ for the cache-assisted
multihop transmission, the\textit{ layer 1 CoMP band} and the \textit{layer
2 CoMP band} with size $W_{c}$ for the layer 1 and layer 2 CoMP transmissions
respectively, where $W_{b}+2W_{c}=W$. As a result, these three transmissions
can occur simultaneously without causing interference to each other.
Furthermore, the multihop bandwidth $W_{b}$ is uniformly divided
into $M$ subbands and each node is allocated with one subband such
that the following condition is satisfied. 

\begin{condition}[Spatial reuse distance]\label{cond:FR}Any two nodes
with distance no more than $r_{I}$ is allocated with different subbands,
where $r_{I}>2r_{\textrm{max}}$ is a system parameter, and $r_{max}$
is defined in Assumption \ref{asm:Place}. 

\end{condition}

The following lemma gives the number of subbands that is required
to satisfy the above condition. 
\begin{lem}
[Admissible frequency reuse factor]\label{lem:FRboundM}There exists
a frequency reuse scheme which has $M\leq\left(\frac{2r_{I}}{r_{\textrm{min}}}+1\right)^{2}+1$
subbands and satisfies Condition \ref{cond:FR}, where $r_{min}$
is the minimum distance between any two nodes as defined in Assumption
\ref{asm:Place}.
\end{lem}

Please refer to Appendix \ref{sub:Proof-of-LemmaCM} for the proof. 

\begin{figure}
\begin{centering}
\textsf{\includegraphics[clip,width=85mm]{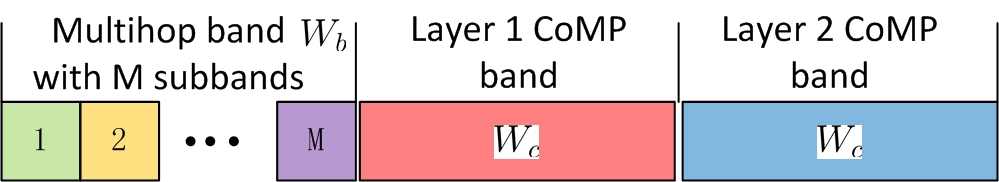}}
\par\end{centering}

\protect\caption{\label{fig:freqplan}{\small{}An illustration of the overall frequency
planning scheme.}}
\end{figure}

The overall frequency planning scheme is illustrated in Fig. \ref{fig:freqplan}.
Note that in mobile adhoc networks, dynamic subband allocation (such
as those studied in \cite{Iacobelli_CIP10_DFA,Elsner_ICUMT10_DFA})
is required for cache-assisted multihop transmission, where each node
exchanges some information with the neighbor nodes and determines
its multihop transmission subband dynamically to avoid strong interference. 

Finally, we adopt uniform power allocation where the power allocated
to each subband is proportional to the bandwidth of the subband. For
example, at each node, the power allocated on the multihop subband
is $\frac{W_{b}P}{W_{b}+MW_{c}}$ and the power allocated on the CoMP
band is $\frac{MW_{c}P}{W_{b}+MW_{c}}$. The total transmit power
of a node is given by $\frac{MW_{c}P}{W_{b}+MW_{c}}+\frac{W_{b}P}{W_{b}+MW_{c}}=P$.

\subsection{Online PHY Mode Determination\label{sub:Physical-Layer-Transmission}}

When node $n$ requests file $l_{n}$, it first determines the PHY
mode. If $q_{l_{n}}<0.5$, node $n$ uses the cache-assisted multihop
transmission to request file $l_{n}$; otherwise, it uses the cache-induced
dual-layer Co-MIMO transmission. The PHY mode is fixed during the
transmission of the entire file $l_{n}$.

\subsection{Online Cache-assisted Multihop Transmission\label{sub:Cache-assisted-Multihop-Transmis}}

The cache-assisted multihop transmission for a node $n$ is summarized
below. \vspace{0.1in}

\textbf{\textit{Cache-assisted Multihop Transmission}} (for node $n$
requesting file $l_{n}$ with multihop cache mode)

\textbf{\small{}Step 1 (Selection of Source Node Set $\mathcal{B}_{n}$
at node $n$)}{\small \par}

\textbf{\small{}\ \ \ \ }\textit{\small{}1a (Request Broadcasting):}{\small{}
Node $n$ broadcasts a REQ message which contains its position and
the requested file index $l_{n}$ to the nodes which are no more than
$r_{\textrm{RB}}=\left(2\sqrt{\left\lceil 1/q_{l_{n}}\right\rceil -1}+1\right)r_{\textrm{max}}$
away from node $n$. Specifically, if the distance between node $n^{'}$
and node $n$ is larger than $r_{\textrm{RB}}$, node $n^{'}$ will
discard the REQ message from node $n$; otherwise, node $n^{'}$ will
1) forward the REQ message to the neighbor nodes and 2) send an ACK
message which contains its position to node $n$.}{\small \par}

\textbf{\small{}\ \ \ \ }\textit{\small{}1b (Source Nodes Selection):}{\small{}
Based on the ACK messages from the nearby nodes, node $n$ chooses
the nearest nodes which have a total number of no less than $\left(1-q_{l_{n}}\right)L_{S}$
parity bits for each segment of the requested file $l_{n}$ as the
source nodes. Specifically, let $r_{n}^{*}=\min\: r,\:\textrm{s.t. }\left|\left\{ n^{'}:\: r_{n,n^{'}}\leq r\right\} \right|\geq\left\lceil 1/q_{l_{n}}\right\rceil $.
Then the set of source nodes for node $n$ is given by $\mathcal{B}_{n}=\left\{ n^{'}\neq n:\: r_{n,n^{'}}\leq r_{n}^{*}\right\} $. }{\small \par}

\textbf{\small{}\ \ \ \ }\textit{\small{}1c (Load Partitioning):}{\small{}
Node $n$ determines the load partitioning among the source nodes.
Specifically, for each requested file segment, node $n$ will obtain
$q_{l_{n}}L_{S}$ parity bits from each node in $\overline{\mathcal{B}}_{n}$
and $\frac{\left(1-q_{l_{n}}-\left|\overline{\mathcal{B}}_{n}\right|q_{l_{n}}\right)L_{S}}{\left|\mathcal{B}_{n}\right|-\left|\overline{\mathcal{B}}_{n}\right|}$
parity bits from each node in $\mathcal{B}_{n}\backslash\overline{\mathcal{B}}_{n}$,
where $\overline{\mathcal{B}}_{n}=\left\{ n^{'}\neq n:\: r_{n,n^{'}}<r_{n}^{*}\right\} $. }{\small \par}

\textbf{\small{}Step 2 (Multihop Routing and Transmission)}{\small \par}

\textbf{\small{}\ \ \ \ }\textit{\small{}2a (Multihop Routing
Path Establishment): }{\small{}For each source node $n^{'}\in\mathcal{B}_{n}$,
node $n$ sends a REQm message to node $n^{'}$ to establish a }\textit{\small{}multihop}{\small{}
}\textit{\small{}routing path}{\small{} between them. The REQm message
contains the positions of node $n^{'}$ and $n$, the requested file
index $l_{n}$, and the number of requested parity bits per segment. }{\small \par}

\textbf{\small{}\ \ \ \ }\textit{\small{}2b (Multihop Transmission):
}{\small{}Each node $n^{'}$ in $\mathcal{B}_{n}$ sends the requested
parity bits as indicated by the REQm message to node $n$ along the
multihop routing path established in step 2a. The source node set
$\mathcal{B}_{n}$ and the multihop routing path are fixed during
the transmission of the entire file $l_{n}$.}\vspace{0.1in}

\begin{figure}
\begin{centering}
\textsf{\includegraphics[clip,width=80mm]{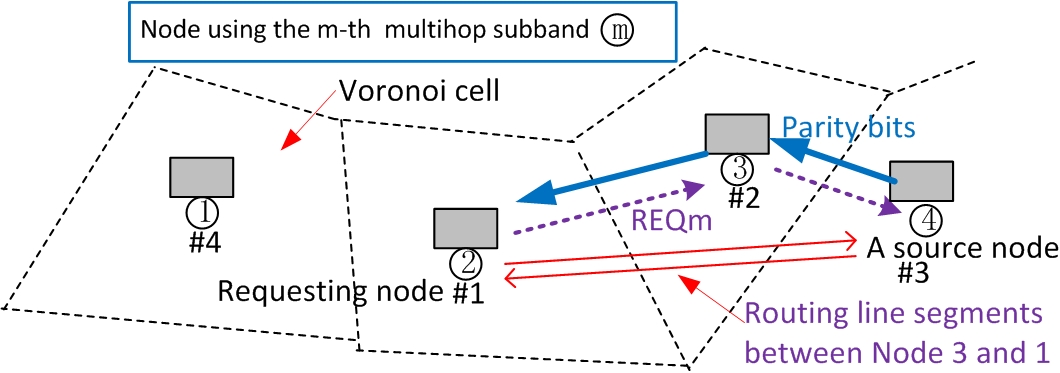}}
\par\end{centering}

\protect\caption{\label{fig:Multihop_Tx}{\small{}An illustration of multihop routing
and transmission, where the requesting node 1 first sends REQm to
a source node $3$ and then source node 3 sends the requested parity
bits to node $1$.}}
\end{figure}

In step 1, the requesting node chooses the nearest nodes containing
$\left(1-q_{l_{n}}\right)L_{S}$ \textit{multihop parity bits} for
each segment of the requested file $l_{n}$ as the source set $\mathcal{B}_{n}$.
In step 2, the requesting node establishes the multihop routing path
(route table) to and from the serving nodes in $\mathcal{B}_{n}$
using a \textit{geometry-based routing}. Specifically, the coverage
area is divided into $N$ Voronoi cells as illustrated in Fig. \ref{fig:Multihop_Tx}.
Then the multihop routing path from node $n$ to node $n^{'}$ consists
of a sequence of hops along a \textit{routing line segment} connecting
node $n$ and node $n^{'}$. In each hop, the REQm message (defined
in step 2a) from node $n$ to node $n^{'}$ are transferred from one
Voronoi cell (node) to another in the order in which they intersect
the routing line segment. For example, in Fig. \ref{fig:Multihop_Tx},
the routing line segment from node 1 to node 3 intersects cell 2 and
cell 3. Then node 1 sends the REQm to node 3 via multihop transmission
over the route ``node 1$\rightarrow$node 2$\rightarrow$node 3''.
The routing path from node $n^{'}$ to $n$ can be obtained by reversing
the routing path from node $n$ to $n^{'}$. Note that each node on
the routing path from node $n$ to node $n^{'}$ can calculate the
node index at the previous and the next hop based on the positions
of node $n$ and node $n^{'}$ in the REQm message, and the positions
of the adjacent nodes. Hence, the route table at each node can be
constructed in a distributed way using the position information in
the REQm message. Finally, the requesting nodes obtain the multihop
parity bits of the requested files from $\mathcal{B}_{n}$ via the
established multihop routes. Together with the $q_{l_{n}}L_{S}$ parity
bits of each segment stored at the local cache, node $n$ can decode
each segment of file $l_{n}$. 

\begin{figure}
\begin{centering}
\textsf{\includegraphics[clip,width=80mm]{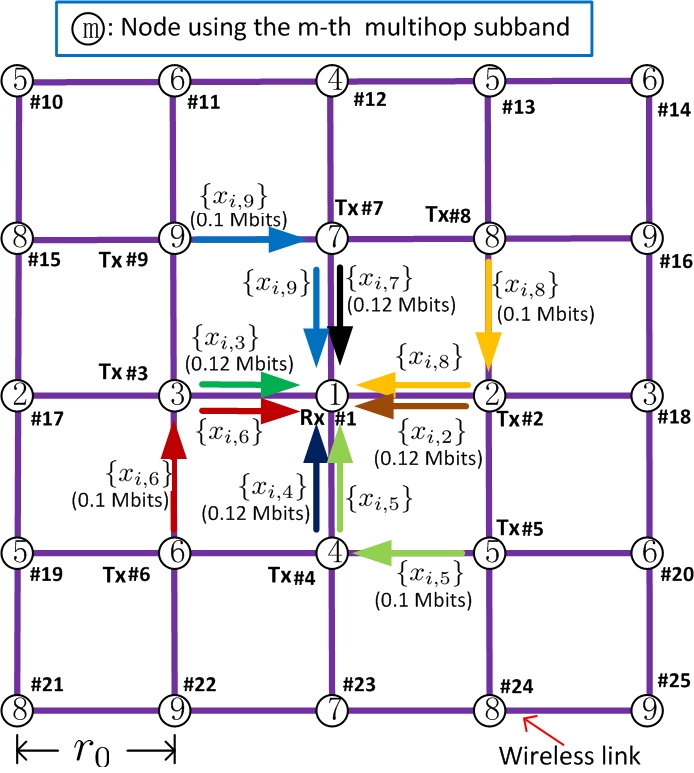}}
\par\end{centering}

\protect\caption{\label{fig:grid-Tx}{\small{}An illustration of cache-assisted multihop
transmission in a regular wireless adhoc network, where node $1$
requests file $1$ associated with multihop cache mode. The $i$-th
segment (with size $L_{S}=1$ Mbit) of file 1 is encoded into $N=25$
parity blocks of 0.12 Mbits and node $n$ caches the $n$-th parity
block $x_{i,n}$ (0.12M parity bits).}}
\end{figure}

Figure 7 illustrates a toy example of the overall cache-assisted multihop
transmission. Node 1 requests file {\small{}1} of segment size $L_{S}=$1
Mbits using multihop mode. The source node set $\mathcal{B}_{1}=\left\{ 2,3,...,9\right\} $
determined by step 1 contains $0.96$M multihop parity bits. After
step 2a, node 1 has established multihop paths to and from the source
nodes $\left\{ 2,3,...,9\right\} $ as illustrated in Fig. \ref{fig:grid-Tx}.
Then node 1 obtains $0.88$M parity bits for every segment of file
1 from nodes in $\mathcal{B}_{1}$ using multihop transmission. Together
with the locally cached 0.12M parity bits for each segment of file
1, node 1 can decode file 1.

\subsection{Online Cache-induced Dual-layer CoMP Transmission\label{sub:Cache-induced-Co-MIMO-Transmissi}}

The cache-induced dual-layer CoMP transmission is illustrated in Fig.
{\small{}\ref{fig:cachePHYmodes}} and is summarized below.\vspace{0.1in}

\textbf{\textit{Cache-induced Dual-layer CoMP Transmission}}

\textbf{\small{}Step 1 (Dual-Layer CoMP Tx Node Clustering):}{\small{}
Each CoMP layer $\mathcal{N}_{C}^{1}$ ($\mathcal{N}_{C}^{2}$) is
further partitioned into CoMP clusters of size $N_{c}$ so that $\mathcal{N}_{C}^{1}=\cup_{j}\mathcal{G}_{1,j}^{\textrm{Tx}}$
and $\mathcal{N}_{C}^{2}=\cup_{j}\mathcal{G}_{2,j}^{\textrm{Tx}}$
. For example, in Fig. \ref{fig:cachePHYmodes}, we have $\mathcal{G}_{1,1}^{\textrm{Tx}}=\left\{ 1,6\right\} $,
$\mathcal{G}_{1,2}^{\textrm{Tx}}=\left\{ 3,8\right\} $, $\mathcal{G}_{2,1}^{\textrm{Tx}}=\left\{ 2,5\right\} $
and $\mathcal{G}_{2,2}^{\textrm{Tx}}=\left\{ 4,7\right\} $.}{\small \par}

\textbf{\small{}Step 2 (Dual-Layer CoMP Rx Node Clustering): }{\small{}Suppose
the requesting node $n\in\mathcal{N}_{C}^{2}$ . Each requesting node
$n$ broadcasts the requested file index $l_{n}$ (associated with
CoMP cache mode) to the nearest nodes in $\mathcal{N}_{C}^{1}$. The
$j$-th CoMP cluster of layer 1 registers the requesting nodes and
let $\mathcal{G}_{1,j}^{\textrm{Rx}}\subset\mathcal{N}_{C}^{2}$ be
the set of all requesting nodes associated with the $j$-th layer
1 cluster $\mathcal{G}_{1,j}^{\textrm{Tx}}$. Fig. \ref{fig:cachePHYmodes}
illustrates an example in which $\mathcal{G}_{1,1}^{\textrm{Rx}}=\left\{ 2,5\right\} $,
$\mathcal{G}_{1,2}^{\textrm{Rx}}=\left\{ 4\right\} $, $\mathcal{G}_{2,1}^{\textrm{Rx}}=\left\{ 6\right\} $
and $\mathcal{G}_{2,2}^{\textrm{Rx}}=\left\{ 3,8\right\} $. }{\small \par}

\textbf{\small{}Step 3 (CoMP transmission in each cluster):}{\small{}
At each time slot, the nodes in $\mathcal{G}_{1,j}^{\textrm{Tx}}$
employ CoMP to jointly transmit the cached parity bits to the nodes
in $\mathcal{G}_{1,j}^{\textrm{Rx}}$ simultaneously. On the other
hand, node $n\in\mathcal{G}_{1,j}^{\textrm{Rx}}$ keeps receiving
the requested portion of parity bits for each segment of file $l_{n}$
($\left(1-q_{l_{n}}\right)L_{S}$ parity bits per segment) until all
the requested segments of file $l_{n}$ is received. The CoMP transmission
in the $j$-th layer 2 cluster is similar.}\vspace{0.1in}

After clustering in step 1 and 2, the nodes in $\mathcal{G}_{1,j}^{\textrm{Tx}}$
and $\mathcal{G}_{1,j}^{\textrm{Rx}}$ forms a \textit{MISO broadcast
channel topology} with $N_{c}$ transmit antennas and $\left|\mathcal{G}_{1,j}^{\textrm{Rx}}\right|$
single receive antenna users as illustrated in Fig. {\small{}\ref{fig:cachePHYmodes}},
where the nodes in $\mathcal{G}_{1,1}^{\textrm{Tx}}=\left\{ 1,6\right\} $
forms a virtual transmitter with two distributed transmit antennas
serving the two single antenna nodes in $\mathcal{G}_{1,1}^{\textrm{Rx}}=\left\{ 2,5\right\} $.
The parity bits of the requested file segments for nodes in $\mathcal{G}_{1,j}^{\textrm{Rx}}$
can be transmitted simultaneously from the nodes in $\mathcal{G}_{1,j}^{\textrm{Tx}}$
using spatial multiplexing and therefore, the network throughput can
be increased. Note that the serving nodes in $\mathcal{G}_{1,j}^{\textrm{Tx}}$
will transmit the same parity bits (codeword) to the same requesting
node in $\mathcal{G}_{1,j}^{\textrm{Rx}}$. Let $\mathcal{C}_{1,j}\left(\left\{ \mathbf{\Sigma}_{n},\sigma_{n}^{2}:n\in\mathcal{G}_{1,j}^{\textrm{Rx}}\right\} \right)$
denote the\textit{ }capacity region of the $j$-th layer 1 MISO BC
(cluster) for a given set of transmit covariance matrices $\left\{ \mathbf{\Sigma}_{n},\forall n\in\mathcal{G}_{1,j}^{\textrm{Rx}}\right\} $,
where $\sigma_{n}^{2}=W_{c}+I_{n}^{c}$ is the variance of the effective
noise at node $n$, and $I_{n}^{c}$ is the inter-cluster interference
seen by node $n$. Then any rate tuple $\left(c_{n},\forall n\in\mathcal{G}_{1,j}^{\textrm{Rx}}\right)\in\mathcal{C}_{1,j}\left(\left\{ \mathbf{\Sigma}_{n},\sigma_{n}^{2}:n\in\mathcal{G}_{1,j}^{\textrm{Rx}}\right\} \right)$
is achievable for the nodes in $\mathcal{G}_{1,j}^{\textrm{Rx}}$.
The expression of $\mathcal{C}_{1,j}\left(\left\{ \mathbf{\Sigma}_{n},\sigma_{n}^{2}:n\in\mathcal{G}_{1,j}^{\textrm{Rx}}\right\} \right)$
is well known (see e.g. \cite{weingarten2006capacity}) and thus is
omitted here for conciseness. Similarly, the $j$-th layer 2 cluster
$\mathcal{G}_{2,j}^{\textrm{Tx}}$ and the associated nodes in $\mathcal{G}_{2,j}^{\textrm{Rx}}$
also form a MISO BC.
\begin{rem}
In practice, it is difficult for each node to have global channel
state information (CSI) for the entire network. Hence, we consider
CoMP clustering so that each node only needs to know the local CSI
within its cluster. In practice, we can allocate a dedicated control
channel for each node to collect the local CSI. The CSI signaling
usually consumes much less bandwidth compared to the data transmission
because the former needs to be done on a per frame basis but the latter
needs to be done on a per-symbol basis. 
\end{rem}

\section{Throughput Analysis in Regular Wireless Adhoc Networks\label{sec:Performance-Analysis}}

In this section, we analyze the per node throughput of the proposed
PHY caching scheme for regular wireless adhoc networks. 
\begin{defn}
[Regular Adhoc Network]In a regular wireless adhoc network, the
$N$ nodes are placed on a grid as illustrated in Fig. \ref{fig:grid-Tx}.
The distance between the adjacent nodes is $r_{0}$.
\end{defn}

Similar to \cite{Gitzenis_TIT13_wirelesscache}, we assume symmetric
traffic model where all nodes have the same throughput requirement
$R$ and all files have the same size, i.e., $F_{l}=F,\forall l$.
To avoid boundary effects, we let $N\rightarrow\infty$. As a comparison,
we will first analyze the performance of a baseline\textit{ multihop
caching scheme} in which the PHY transmission scheme for all nodes
is based on the cache-assisted multihop transmission summarized in
Section \ref{sub:Cache-assisted-Multihop-Transmis}. After deriving
closed form expressions of the per node throughput of both schemes,
we will quantify the benefit of the proposed PHY caching scheme relative
to the multihop caching scheme.

\subsection{Per Node Throughput of Multihop Caching Scheme}

In the multihop caching scheme, we set $r_{I}=2.5r_{0}$. As a result,
the multihop bandwidth $W_{b}=W$ is divided into $M=9$ subbands
as illustrated in Fig. \ref{fig:grid-Tx}. In Fig. \ref{fig:grid-Tx},
we also illustrate the cache-assisted multihop transmission scheme
for a regular wireless adhoc network, where only two adjacent nodes
can form a link and communicate with each other directly. First, we
derive the average rate of each link. 
\begin{lem}
\label{lem:RbA}In the cache-assisted multihop transmission, the average
rate of each link is given by $WR_{b}$ (nats per second), where 
\begin{equation}
R_{b}=\frac{1}{9}\log\left(1+\frac{9Pr_{0}^{-\alpha}}{W+9PI_{R}}\right),\label{eq:RbP}
\end{equation}
\begin{eqnarray*}
I_{R} & = & \sum_{i=1}^{\infty}\frac{1}{r_{0}^{\alpha}}\left[\sum_{j=1}^{\infty}2\left(\left(3i+1\right)^{2}+9j^{2}\right)^{-\frac{\alpha}{2}}+\left|3i+1\right|^{-\alpha}\right]\\
 & + & \sum_{i=1}^{\infty}\frac{1}{r_{0}^{\alpha}}\left[\sum_{j=1}^{\infty}2\left(\left(3i-1\right)^{2}+9j^{2}\right)^{-\frac{\alpha}{2}}+\left|3i-1\right|^{-\alpha}\right]\\
 & + & \sum_{i=1}^{\infty}\frac{2}{r_{0}^{\alpha}}\left(9i^{2}+1\right)^{-\frac{\alpha}{2}}=\Theta\left(1\right).
\end{eqnarray*}

\end{lem}

Please refer to Appendix \ref{sub:Proof-of-LemmaRbA} for the proof
of Lemma \ref{lem:RbA}. From Lemma \ref{lem:RbA}, we obtain the
achievable per node throughput in the following theorem.
\begin{thm}
[Per node throughput of multihop caching]\label{thm:PBTA}Under
the conventional multihop caching scheme, the per node throughput
is 
\[
\Gamma_{B}\left(\mathbf{q}\right)=\frac{WR_{b}}{\sum_{l=1}^{L}p_{l}\psi\left(q_{l}\right)}.
\]
where 
\begin{eqnarray}
\psi\left(q\right) & = & \frac{\phi\left(q\right)\left(1-q\right)-\frac{2}{3}\left(\phi^{3}\left(q\right)-\phi\left(q\right)\right)q}{2},\label{eq:RbA}\\
\phi\left(q\right) & = & \left\lceil \frac{-1+\sqrt{\frac{2}{q}-1}}{2}\right\rceil .\nonumber 
\end{eqnarray}

\end{thm}

The intuition behind Theorem \ref{thm:PBTA} is as follows. As can
be seen in Fig. \ref{fig:grid-Tx}, for each node, the number of nodes
with the nearest distance ($r_{0}$) from it is $4$, that with the
second nearest distance ($\sqrt{2}r_{0}$) is $8$, and that with
the $m$-th nearest distance is $4m$. Let $\mathcal{B}_{n,m}$ denote
the set of nodes with the $m$-th nearest distance from node $n$.
Suppose node $n$ requests the $l$-th file. Then $\phi\left(q_{l}\right)$
is the maximum number of hops between node $n$ and its source nodes
in $\mathcal{B}_{n}$. For example, in Fig. \ref{fig:grid-Tx}, $q_{l}=0.12$
and thus the maximum number of hops between node $1$ and its source
nodes is $\phi\left(0.12\right)=2$. It can be shown that the average
traffic rate induced by a single node is $\sum_{l=1}^{L}p_{l}2\psi\left(q_{l}\right)R$,
where $T_{l}\triangleq2\psi\left(q_{l}\right)R$ is the traffic rate
induced by a single node requesting the $l$-th file. Moreover, since
the ratio between the number of links and the number of nodes is 2
as $N\rightarrow\infty$, the traffic rate on each link is $\sum_{l=1}^{L}p_{l}\psi\left(q_{l}\right)R$.
Clearly, the per node throughput requirement $R$ can be satisfied
if the traffic rate on each link does not exceed the average rate
of each link, i.e., $\sum_{l=1}^{L}p_{l}\psi\left(q_{l}\right)R\leq WR_{b}$.
Hence, the per node throughput is $\Gamma_{B}\left(\mathbf{q}\right)$.
Please refer to Appendix \ref{sub:Proof-of-TheoremPBTA} for the detailed
proof of Theorem \ref{thm:PBTA}.

\subsection{Per Node Throughput of the Proposed PHY Caching Scheme }

Recall that for $M=9$, the bandwidth of a multihop subband is $\frac{W-2W_{c}}{9}$
and the transmit power on this bandwidth is $\frac{\left(W-2W_{c}\right)P}{W+7W_{c}}$.
The noise power is $\frac{W-2W_{c}}{9}$ and the interference power
from the interfering nodes transmitting on the same multihop subband
is $\frac{\left(W-2W_{c}\right)P}{W+7W_{c}}I_{R}$, where $I_{R}$
is given in Lemma 2. Hence the SINR of each link in the cache-assisted
multihop transmission is $\frac{9Pr_{0}^{-\alpha}}{W+7W_{c}+9PI_{R}}$,
and the corresponding rate is given by $\left(W-2W_{c}\right)R_{m}\left(W_{c}\right)$,
where
\begin{equation}
R_{m}\left(W_{c}\right)=\frac{1}{9}\log\left(1+\frac{9Pr_{0}^{-\alpha}}{W+7W_{c}+9PI_{R}}\right).\label{eq:Rm}
\end{equation}
Clearly, $R_{m}\left(W_{c}\right)$ is bounded as $R_{m}^{U}\geq R_{m}\left(W_{c}\right)\geq R_{m}^{L}$,
where $R_{m}^{U}=R_{b}$ and 
\begin{eqnarray*}
R_{m}^{L} & = & \frac{1}{9}\log\left(1+\frac{2Pr_{0}^{-\alpha}}{W+2PI_{R}}\right).
\end{eqnarray*}
When a node requests a file with CoMP cache mode, it is served using
the cache-induced dual-layer CoMP in Section \ref{sub:Cache-induced-Co-MIMO-Transmissi}
and the average rate has no closed-form expression. The following
theorem gives closed-form bounds for the average rate in this case.
\begin{thm}
[Average rate bounds for cache-induced dual-layer CoMP]\label{thm:LoCoMIMObound}Let
\[
G_{C}=\frac{4}{r_{0}^{\alpha}}\sum_{i=1}^{\infty}\sum_{j=1}^{\infty}\left[\left(\frac{\sqrt{2}}{2}+i\right)^{2}+\left(\frac{\sqrt{2}}{2}+j\right)^{2}\right]^{-\frac{\alpha}{2}}=\Theta\left(1\right)
\]
$\rho=\frac{1}{2}\left(1-\frac{r_{0}^{-\alpha}}{G_{C}}\right)^{2}$,
$R_{c}^{U}=\log\left(1+\frac{9PG_{C}}{W}\right)$ and $R_{c}^{L}=\rho\log\left(1+\frac{2Pr_{0}^{-\alpha}}{W}\right)$.
The average rate of a node served using the cache-induced dual-layer
CoMP is $W_{c}R_{c}\left(W_{c}\right)$, where $R_{c}\left(W_{c}\right)$
is bounded as 
\[
R_{c}^{U}\geq R_{c}\left(W_{c}\right)\geq R_{c}^{L}+O\left(PN_{c}^{-\frac{\alpha-2}{2\left(\alpha-1\right)}}\right).
\]

\end{thm}

In Theorem \ref{thm:LoCoMIMObound}, the upper bound is obtained using
the cut set bound between all nodes in $\mathcal{N}_{C}^{1}$ and
a node in $\mathcal{N}_{C}^{2}$. The lower bound is obtained using
an achievable scheme which ensures that the inter-cluster interference
is negligible ($O\left(PN_{c}^{\frac{2-\alpha}{2\left(\alpha-1\right)}}\right)$
) for large CoMP cluster size $N_{c}$. Then we can focus on studying
the achievable sum rate of the MISO BC within each cluster. Finally,
the lower bound can be derived from the achievable sum rate and the
small term $O\left(PN_{c}^{-\frac{\alpha-2}{2\left(\alpha-1\right)}}\right)$
in the lower bound is due to the inter-cluster interference of order
$O\left(PN_{c}^{\frac{2-\alpha}{2\left(\alpha-1\right)}}\right)$.
Please refer to Appendix \ref{sub:Proof-of-TheoremCoMIMO} for the
detailed proof. 
\begin{cor}
[Per node throughput bounds of PHY caching]\label{cor:PBA}Under
the proposed PHY caching scheme, the per node throughput $\Gamma_{A}\left(\mathbf{q}\right)$
is given by
\begin{equation}
\Gamma_{A}\left(\mathbf{q}\right)=\frac{WR_{m}\left(W_{c}^{*}\right)}{Q_{\Omega\left(\mathbf{q}\right)}+2Q_{\overline{\Omega}\left(\mathbf{q}\right)}R_{m}\left(W_{c}^{*}\right)/R_{c}\left(W_{c}^{*}\right)},\label{eq:TAqgen}
\end{equation}
where $Q_{\Omega\left(\mathbf{q}\right)}=\sum_{l\in\Omega\left(\mathbf{q}\right)}p_{l}\psi\left(q_{l}\right)$,
$Q_{\overline{\Omega}\left(\mathbf{q}\right)}=\sum_{l\in\overline{\Omega}\left(\mathbf{q}\right)}p_{l}\left(1-q_{l}\right)$,
and $W_{c}^{*}$ is the unique solution of 
\begin{equation}
Q_{\overline{\Omega}\left(\mathbf{q}\right)}\left(W-2W_{c}\right)R_{m}\left(W_{c}\right)=Q_{\Omega\left(\mathbf{q}\right)}W_{c}R_{c}\left(W_{c}\right).\label{eq:Wstar}
\end{equation}
Moreover, $\Gamma_{A}\left(\mathbf{q}\right)$ is bounded as $\Gamma_{A}^{U}\left(\mathbf{q}\right)\geq\Gamma_{A}\left(\mathbf{q}\right)\geq\Gamma_{A}^{L}\left(\mathbf{q}\right)+O\left(PN_{c}^{-\frac{\alpha-2}{2\left(\alpha-1\right)}}\right)$
with 
\begin{equation}
\Gamma_{A}^{a}\left(\mathbf{q}\right)=\frac{WR_{m}^{a}}{Q_{\Omega\left(\mathbf{q}\right)}+2Q_{\overline{\Omega}\left(\mathbf{q}\right)}R_{m}^{a}/R_{c}^{a}},\label{eq:Taqgen}
\end{equation}
for $a\in\left\{ L,U\right\} $. Finally, as $P,N_{c}\rightarrow\infty$
such that $PN_{c}^{-\frac{\alpha-2}{2\left(\alpha-1\right)}}\rightarrow0$,
we have 
\begin{equation}
\Gamma_{A}\left(\mathbf{q}\right)\rightarrow\frac{W}{9Q_{\Omega\left(\mathbf{q}\right)}}\log\left(1+\frac{r_{0}^{-\alpha}}{I_{R}}\right).\label{eq:TAq}
\end{equation}

\end{cor}

Please refer to Appendix \ref{sub:Proof-of-CorollaryPBA} for the
proof. Note that the upper and lower bounds $\Gamma_{A}^{U}\left(\mathbf{q}\right),\Gamma_{A}^{L}\left(\mathbf{q}\right)$
are asymptotically tight at high SNR.

\subsection{Order Optimal Cache Content Replication Solution}

The per node throughput of both PHY caching and multihop caching is
a non-concave function of the cache content replication vector $\mathbf{q}$.
To make the analysis tractable, we aim at finding the \textit{order-optimal}
cache content replication solution. For convenience, the notation
$N,L\overset{\xi}{\rightarrow}\infty$ refers to $N\rightarrow\infty$
and $\lim_{N\rightarrow\infty}\frac{L}{N}=\xi$. Since we assume $NB_{C}>LF$,
we have $\xi\in\left[0,\frac{B_{C}}{F}\right)$. Note that $\xi$
is allowed to be zero. Hence as $N\rightarrow\infty$, $L$ can be
either $\Theta\left(1\right)$ or go to infinity at an order no larger
than $N$. The following corollary follows from Theorem \ref{thm:PBTA}
and Corollary \ref{cor:PBA}. The detailed proof can be found in Appendix
\ref{sub:Proof-of-CorollaryOthpR}.
\begin{cor}
[Per node throughput order in regular networks]\label{cor:OthpR}As
$N,L\overset{\xi}{\rightarrow}\infty$, the per-node throughputs of
the PHY caching and multihop caching schemes satisfy $\Gamma_{A}\left(\mathbf{q}\right)=\Theta\left(\frac{1}{\sum_{l=1}^{L}p_{l}\sqrt{\frac{1}{q_{l}}}}\right)$
and $\Gamma_{B}\left(\mathbf{q}\right)=\Theta\left(\frac{1}{\sum_{l=1}^{L}p_{l}\sqrt{\frac{1}{q_{l}}}}\right)$
respectively.
\end{cor}

Consider the problem of maximizing the order of per node throughput
in Corollary \ref{cor:OthpR}: 
\begin{equation}
\min_{\mathbf{q}}\sum_{l=1}^{L}p_{l}\sqrt{\frac{1}{q_{l}}},\: s.t.\: q_{l}\in\left[\frac{1}{N},1\right],\forall l,\:\sum_{l=1}^{L}q_{l}\leq\frac{B_{C}}{F}.\label{eq:optq}
\end{equation}
The constraint $q_{l}\geq\frac{1}{N}$ is to ensure that there is
at least one complete copy of the $l$-th file in the caches of the
entire network. Problem (\ref{eq:optq}) is convex and the optimal
solution can be easily obtained using numerical method. To facilitate
performance analysis, we characterize the \textit{order-optimal} $\mathbf{q}$
in the following theorem. A cache content replication vector $\mathbf{q}$
is \textit{order-optimal} for problem (\ref{eq:optq}) if the achieved
objective value is on the same order as the optimal objective value
as $N,L\overset{\xi}{\rightarrow}\infty$. Note that the order-optimality
here is w.r.t. problem (\ref{eq:optq}) under the proposed scheme.
It is not the order-optimality in\textbf{ }information theoretic sense.
\begin{thm}
[Order optimal cache content replication]\label{thm:Order-optimal-pR}As
$N,L\overset{\xi}{\rightarrow}\infty$, an order-optimal $\mathbf{q}$
is given by
\begin{equation}
q_{l}^{*}=\min\left(\left(\frac{B_{C}}{F}-\frac{L}{N}\right)\frac{p_{l}^{2/3}}{\sum_{l=1}^{L}p_{l}^{2/3}}+\frac{1}{N},1\right),\forall l.\label{eq:odroptq}
\end{equation}

\end{thm}

Please refer to Appendix \ref{sub:Proof-of-Theorem-OdpR} for the
detailed proof. Theorem \ref{thm:Order-optimal-pR} implies that the
order-optimal cache content replication variable $q_{l}$ is proportional
to $p_{l}^{2/3}$, indicating that a larger portion of the cache capacity
should be allocated to more popular content%
\footnote{This result is consistent with \cite{Jin_ACMISM05_cacheWCDN,Gitzenis_TIT13_wirelesscache}.%
}.

\subsection{Benefits of PHY Caching\label{sub:Cache-induced-MIMO-Cooperation}}

The PHY caching gain is defined as the throughput gap between the
PHY caching and multihop caching. We analyze this gain under the Zipf
content popularity distribution: 
\begin{equation}
p_{l}=\frac{1}{Z_{\tau}\left(L\right)}l^{-\tau},l=1,...,L,\label{eq:zipp}
\end{equation}
where the parameter $\tau$ determines the rate of popularity decline
as $l$ increases, and $Z_{\tau}\left(L\right)=\sum_{l=1}^{L}l^{-\tau}$
is a normalization factor. The Zipf distribution is widely used to
model the Internet traffic \cite{Breslau_INFOCOM99_ZipfLaw,Yamakami_PDCAT06_Zipflaw}.
The Zipf parameter $\tau$ usually ranges from 0.5 to 3 \cite{Yamakami_PDCAT06_Zipflaw}
depending on the application. Higher values of $\tau$ are usually
observed in mobile applications \cite{Yamakami_PDCAT06_Zipflaw}.
Using Theorem \ref{thm:PBTA} and Corollary \ref{cor:PBA}, we can
bound the PHY caching gain $\triangle\Gamma\triangleq\Gamma_{A}\left(\mathbf{q}^{*}\right)-\Gamma_{B}\left(\mathbf{q}^{*}\right)$
under the order-optimal cache content replication $\mathbf{q}^{*}$
in (\ref{eq:odroptq}). 
\begin{cor}
[PHY caching gain]\label{cor:Cache-induced-MIMO-cooperation}The
PHY caching gain is bounded as $\triangle\Gamma_{U}\geq\triangle\Gamma\geq\triangle\Gamma_{L}+O\left(PN_{c}^{-\frac{\alpha-2}{2\left(\alpha-1\right)}}\right)$
with 
\begin{eqnarray}
\triangle\Gamma_{a} & = & \Gamma_{A}^{a}\left(\mathbf{q}^{*}\right)-\Gamma_{B}\left(\mathbf{q}^{*}\right),\label{eq:Taapprx}
\end{eqnarray}
for $a\in\left\{ L,U\right\} $. Moreover, as $P,N_{c}\rightarrow\infty$
such that $PN_{c}^{-\frac{\alpha-2}{2\left(\alpha-1\right)}}\rightarrow0$,
we have $\triangle\Gamma\rightarrow\overline{\triangle\Gamma}$, where
\begin{eqnarray}
\overline{\triangle\Gamma} & = & \frac{W}{9}\log\left(1+\frac{r_{0}^{-\alpha}}{I_{R}}\right)\left(\frac{\sum_{l\in\overline{\Omega}\left(\mathbf{q}^{*}\right)}p_{l}\psi\left(q_{l}^{*}\right)}{Q_{\Omega\left(\mathbf{q}^{*}\right)}\sum_{l=1}^{L}p_{l}\psi\left(q_{l}^{*}\right)}\right)\label{eq:dltThSNR}\\
 & = & \Theta\left(\frac{\frac{B_{C}}{F}\sum_{l\in\overline{\Omega}\left(\mathbf{q}^{*}\right)}\left(1-q_{l}^{*}\right)}{\left[\sum_{l=1}^{L}p_{l}^{2/3}\right]^{3}}\right),\nonumber 
\end{eqnarray}
where $q_{l}^{*}$ is given in (\ref{eq:odroptq}).
\end{cor}

\begin{figure}
\begin{centering}
\includegraphics[width=80mm]{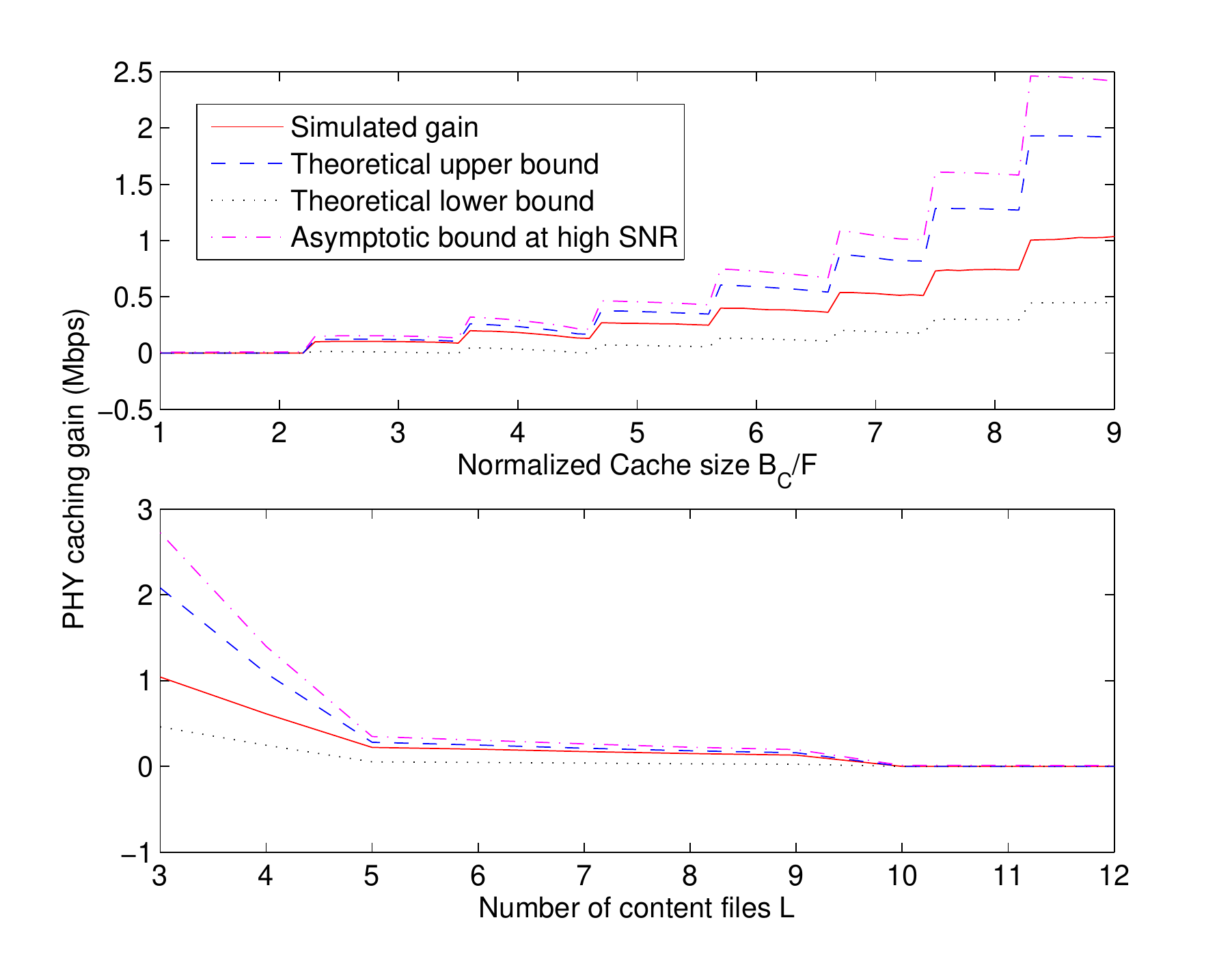}
\par\end{centering}

\protect\caption{\label{fig:VerifyUL}{\small{}Impact of system parameters on the PHY
caching gain in a regular wireless adhoc network with $N=512$ nodes.
The system bandwidth is 1MHz and $\frac{Pr_{0}^{-\alpha}}{W}=10$}dB{\small{}.
The content popularity skewness $\tau=$1. The CoMP cluster size is
$N_{c}=9$. In the upper subplot, $L=12$. In the lower subplot, $B_{C}/F=2$.}}
\end{figure}

According to Corollary \ref{cor:Cache-induced-MIMO-cooperation},
the PHY caching gain can be well approximated by $\overline{\triangle\Gamma}$
at high SNR. Clearly, we have $\overline{\triangle\Gamma}\geq0$.
$\overline{\triangle\Gamma}$ captures the key features of the actual
(simulated) PHY caching gain even at moderate SNR as illustrated in
Fig. \ref{fig:VerifyUL}. From (\ref{eq:dltThSNR}), we have the following
observation about the impact of system parameters on $\overline{\triangle\Gamma}$. 

\textbf{Impact of the normalized cache size $\frac{B_{C}}{F}$:} When
$\frac{B_{C}}{F}<\left(0.5-\frac{1}{N}\right)\sum_{l=1}^{L}l^{-\frac{2}{3}\tau}+\frac{L}{N}$,
we have $q_{l}^{*}<0.5,\forall l$ and $\overline{\triangle\Gamma}=0$.
When $\frac{B_{C}}{F}=\left(0.5-\frac{1}{N}\right)\sum_{l=1}^{L}l^{-\frac{2}{3}\tau}+\frac{L}{N}$,
$\overline{\triangle\Gamma}$ jumps from 0 to a positive value. As
$\frac{B_{C}}{F}$ increases over the region $\left[\left(0.5-\frac{1}{N}\right)\sum_{l=1}^{L}l^{-\frac{2}{3}\tau}+\frac{L}{N},2^{\frac{2}{3}\tau}\left(0.5-\frac{1}{N}\right)\sum_{l=1}^{L}l^{-\frac{2}{3}\tau}+\frac{L}{N}\right)$,
$\overline{\triangle\Gamma}$ keeps positive but its value may fluctuate.
When $\frac{B_{C}}{F}=2^{\frac{2}{3}\tau}\left(0.5-\frac{1}{N}\right)\sum_{l=1}^{L}l^{-\frac{2}{3}\tau}+\frac{L}{N}$,
$\overline{\triangle\Gamma}$ jumps to a larger value. As $\frac{B_{C}}{F}$
continues to increase, $\overline{\triangle\Gamma}$ repeats the pattern
of fluctuations followed by a positive jump. Overall, $\overline{\triangle\Gamma}$
increases with $\frac{B_{C}}{F}$ as shown in the upper subplot of
Fig. \ref{fig:VerifyUL}. 

\textbf{Impact of the number of content files $L$: }In the lower
subplot of Fig. \ref{fig:VerifyUL}, we plot $\overline{\triangle\Gamma}$
versus $L$ when fixing other parameters. It can be observed that
$\overline{\triangle\Gamma}$ decreases with $L$. When $\tau\leq\frac{3}{2}$,
there exists a large enough $L$, such that $\frac{B_{C}}{F}<\left(0.5-\frac{1}{N}\right)\sum_{l=1}^{L}l^{-\frac{2}{3}\tau}+\frac{L}{N}$
and $\overline{\triangle\Gamma}=0$. 

\textbf{Impact of the Content Popularity Skewness $\tau$:} When $\tau>\frac{3}{2}$,
$\sum_{l=1}^{L}p_{l}^{2/3}=\Theta\left(\sum_{l=1}^{L}l^{-\frac{2}{3}\tau}\right)$
is bounded and we can achieve a PHY caching gain of $\Theta\left(1\right)$
even when $L\rightarrow\infty$. On the other hand, when $\tau\leq\frac{3}{2}$,
the normalized cache size $\frac{B_{C}}{F}$ has to increase with
$L$ at the same order as $L\rightarrow\infty$ in order to achieve
a PHY caching gain of $\Theta\left(1\right)$.

\section{Throughput Scaling Laws in General Wireless Adhoc Networks\label{sec:Scaling-Laws-of}}

In this section, we study throughput scaling laws for general wireless
adhoc networks (i.e., the cache-assisted wireless adhoc networks described
in Section \ref{sec:System-Model} employing the proposed PHY caching
scheme) as $N,L\overset{\xi}{\rightarrow}\infty$. We first give the
order of the per node throughput of the proposed PHY caching scheme
with fixed cache content replication $\mathbf{q}$.
\begin{thm}
[Per node throughput order in general networks]\label{thm:CLB}In
a general cache-assisted wireless adhoc network with $N$ nodes, a
per node throughput 
\begin{equation}
R=\Theta\left(\frac{W}{M\sum_{l=1}^{L}p_{l}\sqrt{\frac{1}{q_{l}}}}\log\left(1+\frac{3MP\left(2r_{\textrm{max}}\right)^{-\alpha}}{\left(M+1\right)W+3MI_{A}}\right)\right)\label{eq:equR}
\end{equation}
can be achieved by the proposed PHY caching scheme with fixed cache
content replication $\mathbf{q}$, where $M=\left(\frac{2r_{I}}{r_{\textrm{min}}}+1\right)^{2}+1$
and 
\[
I_{A}=\frac{4P\left(r_{I}-2r_{\textrm{max}}+r_{\textrm{min}}\right)}{r_{\textrm{min}}^{2}\left(r_{I}-2r_{\textrm{max}}\right)^{\alpha-1}}\left(\frac{2}{\alpha-2}+\frac{1}{\alpha-1}+3\right).
\]

\end{thm}

Please refer to Appendix \ref{sub:Proof-of-TheoremCLB} for the proof. 

The factor $\sum_{l=1}^{L}p_{l}\sqrt{\frac{1}{q_{l}}}$ in (\ref{eq:equR})
is due to multihop transmission and it determines the order of per
node throughput. When $q_{l}$ increases, the average number of hops
from the source nodes to the destination nodes decreases and thus
the order of per node throughput increases. Note that the order of
the per node throughput in Theorem \ref{thm:CLB} is consistent with
that of the regular wireless adhoc network in Corollary \ref{cor:OthpR}.
Hence, the order optimal cache content replication solution in general
wireless adhoc networks is also given by (\ref{eq:odroptq}) in Theorem
\ref{thm:Order-optimal-pR} and the following scaling laws follow
straightforward from Theorem \ref{thm:Order-optimal-pR} and \ref{thm:CLB}.
\begin{cor}
[Asymptotic scaling laws of per node throughput]\label{cor:Asymptotic-scaling-laws}Under
the Zipf content popularity distribution in (\ref{eq:zipp}), the
order optimal per node throughput in general wireless adhoc networks
is given by $R^{*}=\Theta\left(\left[\sum_{l=1}^{L}p_{l}^{2/3}\right]^{-3/2}\right)$.
Moreover, when $L=\Theta\left(1\right)$, we have $R^{*}=\Theta\left(1\right)$.
When $N,L\rightarrow\infty$ and $\lim_{N\rightarrow\infty}\frac{L}{N}\in\left[0,\frac{B_{C}}{F}\right)$,
we have
\begin{enumerate}
\item If $0\leq\tau<1$, $R^{*}=\Theta\left(1/\sqrt{L}\right)$.
\item If $\tau=1$, $R^{*}=\Theta\left(\log L/\sqrt{L}\right)$.
\item If $1<\tau<3/2$, $R^{*}=L^{\tau-3/2}$.
\item If $\tau=3/2$, $R^{*}=\log^{-3/2}L$.
\item If $\tau>3/2$, $R^{*}=\Theta\left(1\right)$.
\end{enumerate}
\end{cor}

Using the proposed MDS cache encoding scheme, the order of per node
throughput is significantly improved compared to the Gupta\textendash Kumar
law $\Theta\left(1/\sqrt{N}\right)$ in \cite{GuptaKumar}. This order-wise
throughput gain is referred to as the \textit{cache-assisted multihop
gain}.

\textbf{Impact of the number of content files $L$:} When $L=\Theta\left(1\right)$,
the per node throughput is $\Theta\left(1\right)$ and PHY caching
achieves order gains. When $L\rightarrow\infty$ and $B_{C}=\Theta\left(F\right)$,
the cache-assisted multihop gain depends heavily on the \textit{content
popularity skewness} represented by the parameter $\tau$. Another
interesting case when $L\rightarrow\infty$ and $B_{C}=\Theta\left(L\right)$
is not studied in this paper because we assume $B_{C}=\Theta\left(F\right)$
(in practice, this is usually true since the cache size at each node
is limited). However, it can be shown that in this case, uniform caching
(i.e., $q_{l}=\frac{B_{C}}{LF},l=1,...,L$) is sufficient to achieve
the order optimal per node throughput: $R^{*}=\Theta\left(\frac{1}{\sum_{l=1}^{L}p_{l}\sqrt{\frac{1}{q_{l}}}}\right)=\Theta\left(\sqrt{\frac{B_{C}}{LF}}\right)=\Theta\left(1\right)$,
which still provides order gains compared with the network without
cache. 

\textbf{Impact of the popularity skewness $\tau$:} For a larger $\tau$,
the requests will concentrate more on a few content files and thus
a larger cache-assisted multihop gain can be achieved. When $L\rightarrow\infty$
and $B_{C}=\Theta\left(F\right)$, there are two \textit{critical
popularity skewness} points: $\tau=1$ and $\tau=3/2$. When $\tau>3/2$,
PHY caching can achieve a per node throughput of $\Theta\left(1\right)$
(order gains) even if $B_{C}\ll LF$. When $\tau<1$, if $L=\Theta\left(N\right)$,
PHY caching does not provide order gain, and the per node throughput
scales according to the Gupta\textendash Kumar law $\Theta\left(1/\sqrt{N}\right)$.
When $L$ scales slower than $N$, there is still an order improvement
over the Gupta\textendash Kumar law.
\begin{rem}
Although we focus on the phase fading channel model in Assumption
\ref{asm:channel}, the main results can be extended to more complicated
fading channel models such as Rayleigh fading. This is because the
order of per node throughput in (\ref{eq:equR}) does not depend on
the fading channel model or channel parameters, but only depends on
the content popularity distribution $\mathbf{p}$ and the cache content
replication vector $\mathbf{q}$. For example, under Rayleigh fading
channel, we can still prove that $R=\Theta\left(\frac{1}{\sum_{l=1}^{L}p_{l}\sqrt{\frac{1}{q_{l}}}}\right)$.
As a result, the order optimal cache content replication vector in
Theorem \ref{thm:Order-optimal-pR} does not depend on the channel
parameters, and the scaling laws in Corollary \ref{cor:Asymptotic-scaling-laws}
still hold. However, the exact per node throughput depends on the
channel parameters as shown in Fig. \ref{fig:GainBC}. 
\end{rem}

\section{Numerical Results\label{sec:Numerical-Results-and}}

Consider a general wireless adhoc network with 256 nodes. The locations
of the nodes are randomly generated according to Assumption \ref{asm:Place}
with $r_{0}=100$m, $r_{\textrm{min}}=50$m and $r_{\textrm{max}}=$75m.
The system bandwidth is 1MHz. The transmit power $P$ is chosen such
that $\frac{Pr_{0}^{-\alpha}}{W}=$10dB. The size of each file is
1GB. We assume Zipf popularity distribution with different values
of $L$ and $\tau$. The cluster size in cache-induced dual-layer
CoMP is $N_{c}=9$. We illustrate the gains of PHY caching by comparing
it to multihop caching and the following baselines.
\begin{itemize}
\item \textbf{Baseline 1 (Classical Multihop }\cite{GuptaKumar}\textbf{)}:
This is the classical multihop scheme in \cite{GuptaKumar}.
\item \textbf{Baseline 2 (JCRD }\cite{Gitzenis_TIT13_wirelesscache}\textbf{)}:
The JCRD scheme in \cite{Gitzenis_TIT13_wirelesscache} for P2P systems
or CDN.
\item \textbf{Baseline 3} \textbf{(GreedyDual }\cite{Jin_ACMISM05_cacheWCDN}\textbf{):
}This is the modified GreedyDual caching algorithm in \cite{Jin_ACMISM05_cacheWCDN}.
\end{itemize}

\begin{figure}
\begin{centering}
\includegraphics[width=80mm]{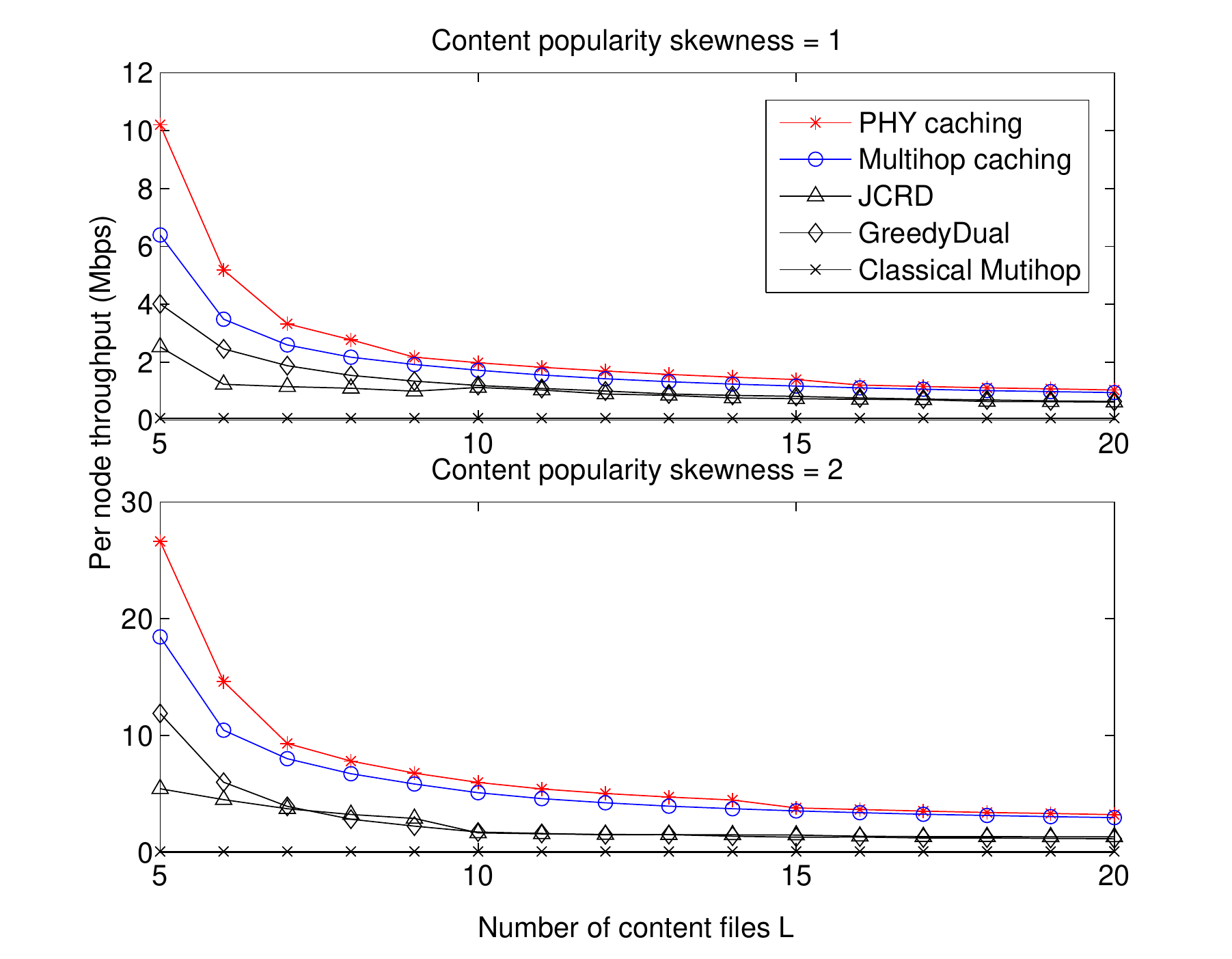}
\par\end{centering}

\protect\caption{\label{fig:GainL}{\small{}Per node throughput versus the number of
content files $L$, where $B_{C}=$4 GB and $\alpha=3.9$.}}
\end{figure}

\begin{figure}
\begin{centering}
\includegraphics[width=80mm]{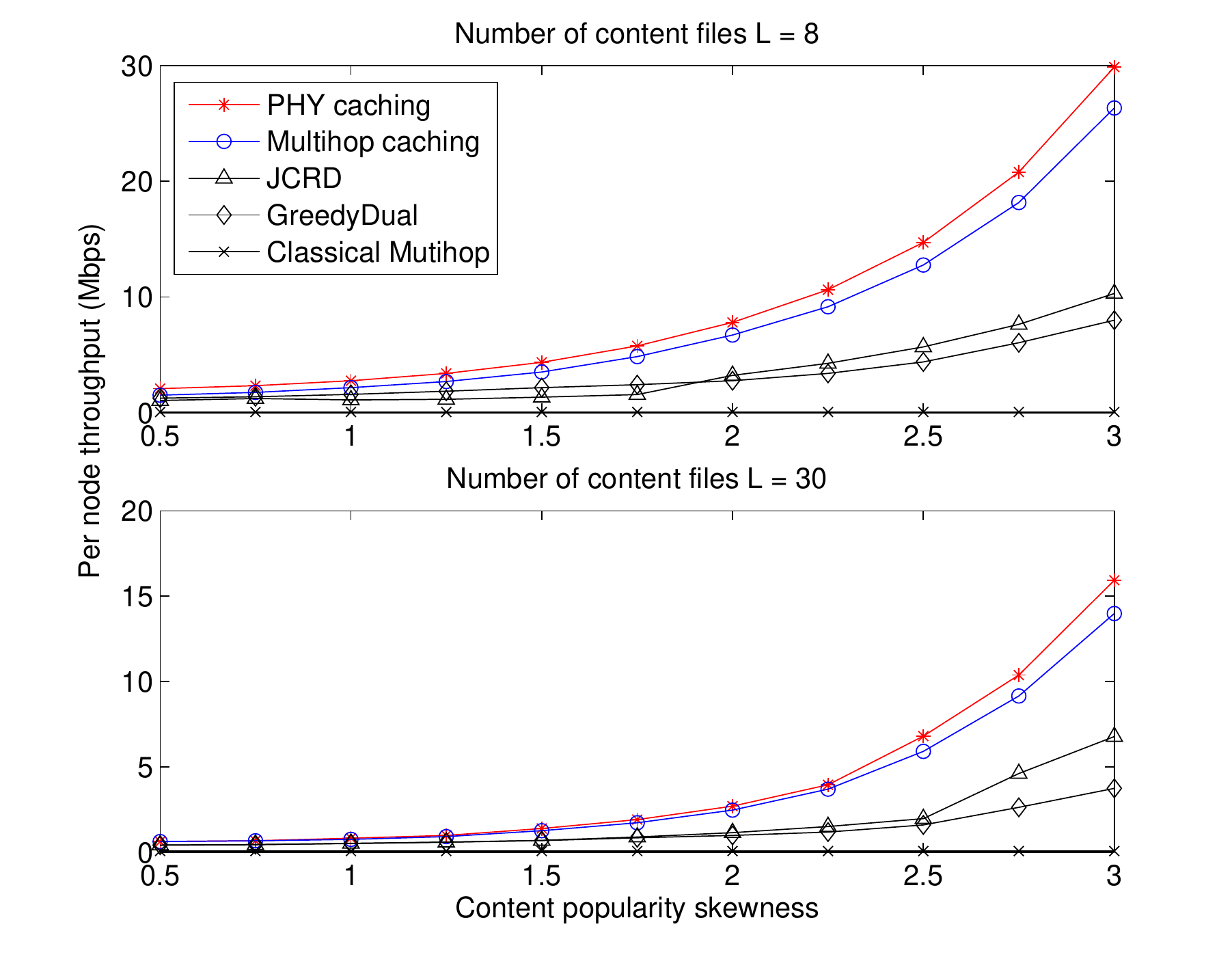}
\par\end{centering}

\protect\caption{\label{fig:Gaintau}{\small{}Per node throughput versus the content
popularity skewness $\tau$, where $B_{C}=$4 GB and $\alpha=3.9$.}}
\end{figure}

\begin{figure}
\begin{centering}
\includegraphics[width=80mm]{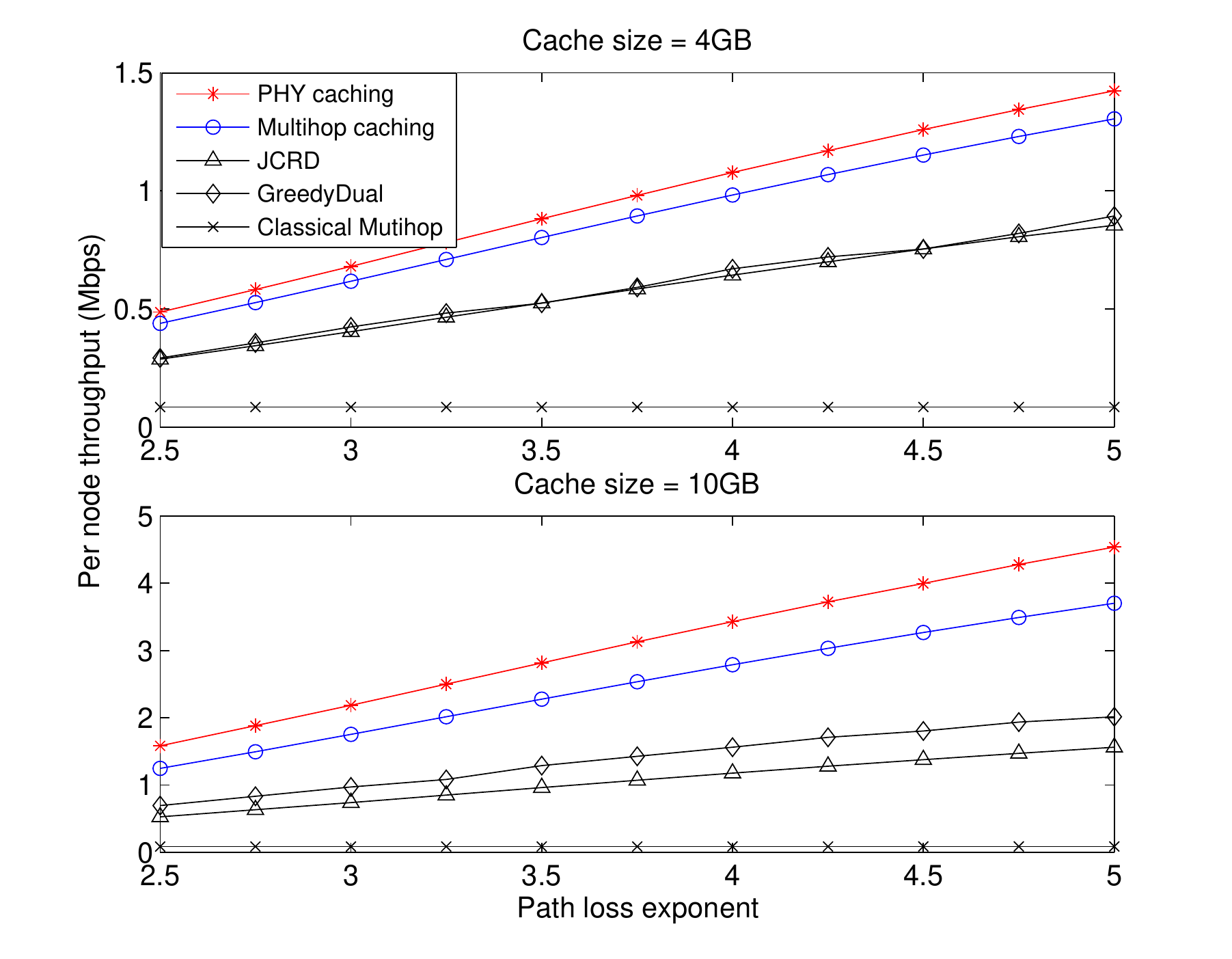}
\par\end{centering}

\protect\caption{\label{fig:GainBC}{\small{}Per node throughput versus the path loss
exponent $\alpha$, where $\tau=1$ and $L=20$.}}
\end{figure}

Fig. \ref{fig:GainL} plots the per node throughput versus the number
of files $L$ for $\tau=1$ and $\tau=2$. The multihop caching scheme
achieves a much higher throughput than the classical multihop scheme
due to the cache-assisted multihopping gain. It also achieves a much
higher throughput than the JCRD and GreedyDual schemes due to the
\textit{MDS encoding gain}. Finally, the proposed PHY caching achieves
the best performance because it can simultaneously exploit the cache-assisted
multihopping gain, the MDS encoding gain and the cache-induced dual-layer
CoMP gain.

In Fig. \ref{fig:Gaintau}, we plot the per node throughput versus
the content popularity skewness $\tau$ for $L=8$ and $L=30$ respectively.
Again, the proposed PHY caching achieves a significant gain over all
baselines. Moreover, even when the cache size at each node is much
smaller than the total content size, it is still possible to achieve
significant PHY caching gains as $\tau$ becomes large. In practice,
the popularity skewness $\tau$ can be large especially for mobile
applications \cite{Yamakami_PDCAT06_Zipflaw}. Hence the PHY caching
can be a very effective way of enhancing the capacity of wireless
networks.

Fig. \ref{fig:GainBC} plots the per node throughput versus the path
loss exponent $\alpha$ for $B_{C}=4$ GB and $B_{C}=10$ GB respectively.
It can be seen that the three PHY caching gains increase with cache
size $B_{C}$.

\section{Conclusion\label{sec:Conclusion}}

In this paper, we propose a PHY caching scheme to enhance the capacity
of wireless adhoc networks. Unlike the existing caching schemes where
the cache scheme is usually independent of the PHY, the proposed PHY
caching may change the PHY topology (from interference topology to
broadcast topology by cache-induced dual-layer CoMP). We establish
the corresponding throughput scaling laws under different regimes
of file popularity, cache and content size. We further study the impact
of various system parameters on the PHY caching gain and provide fundamental
design insight for cache-assisted wireless adhoc network. An interesting
future work is to establish the information theoretical capacity scaling
laws in cache-assisted wireless adhoc networks. 

\appendix

\subsection{Proof of Lemma \ref{lem:FRboundM}\label{sub:Proof-of-LemmaCM}}
\begin{IEEEproof}
Construct a graph where the vertices are the nodes. There is an edge
between any two nodes with distance no more than $r_{I}$. Then finding
a frequency reuse scheme which satisfies Condition \ref{cond:FR}
is a vertex coloring problem. It is known that a graph of degree no
more than $D_{G}$ can have its vertices colored by no more than $D_{G}+1$
colors, with no two neighboring vertices having the same color \cite{Bondy_Elsevier76_GraphTheory}.
One can therefore allocate the cells with no more than $D_{G}+1$
subbands such that Condition \ref{cond:FR} is satisfied. The rest
is to bound $D_{G}$, which is the number of nodes in a circle with
radius $r_{I}$ (including the boundary). Using Assumption \ref{asm:Place}-1),
we have $\frac{D_{G}\pi r_{\textrm{min}}^{2}}{4}\leq\pi\left(r_{I}+\frac{r_{\textrm{min}}}{2}\right)^{2}$,
which implies that $D_{G}\leq\left(\frac{2r_{I}}{r_{\textrm{min}}}+1\right)^{2}$. 
\end{IEEEproof}

\subsection{Proof of Lemma \ref{lem:RbA}\label{sub:Proof-of-LemmaRbA} }
\begin{IEEEproof}
For each link, the bandwidth is $\frac{W}{9}$ and the transmit power
on this bandwidth is $P$. The noise power is $\frac{W}{9}$ and it
can be shown that the interference power is $PI_{R}$. Hence the SINR
of each link is $\frac{9Pr_{0}^{-\alpha}}{W+9PI_{R}}$ and the corresponding
rate is $WR_{b}$. Finally, it follows from $\alpha>2$ that $I_{R}=\Theta\left(1\right)$.
\end{IEEEproof}

\subsection{Proof of Theorem \ref{thm:PBTA}\label{sub:Proof-of-TheoremPBTA}}
\begin{IEEEproof}
Suppose node $n$ requests file $l$. According to the proposed source
node set selection scheme, for each requested file segment, $q_{l}L_{S}$
parity bits are obtained from the local cache at node $n$, a total
number of $4mq_{l}L_{S}$ parity bits are obtained from the source
nodes in $\mathcal{B}_{n,m}$ for $1\leq m<\phi\left(q_{l}\right)$,
and a total number of $\left(1-\left(1+2\phi^{2}\left(q_{l}\right)-2\phi\left(q_{l}\right)\right)q_{l}\right)L_{S}$
parity bits are obtained from the source nodes in $\mathcal{B}_{n,\phi\left(q_{l}\right)}$.
As a result, the traffic rate $T_{l}$ induced by a single node requesting
the $l$-th file is
\begin{eqnarray*}
T_{l} & = & \sum_{m=1}^{\phi\left(q_{l}\right)-1}4m^{2}q_{l}R+\phi\left(q_{l}\right)\left(1-\left(1+2\phi^{2}\left(q_{l}\right)-2\phi\left(q_{l}\right)\right)q_{l}\right)R\\
 & = & \left(\phi\left(q_{l}\right)\left(1-q_{l}\right)-\frac{2}{3}(\phi^{3}\left(q_{l}\right)-\phi\left(q_{l}\right))q_{l}\right)R.
\end{eqnarray*}
Note that the ratio between the number of links and the number of
nodes is $\lim_{N\rightarrow\infty}\frac{2N-2\sqrt{N}}{N}=2$. Since
all links are symmetric, the total traffic rate induced by all nodes
are equally partitioned among all links. Hence, the corresponding
traffic rate on each link is $\frac{1}{2}\sum_{l=1}^{L}p_{l}T_{l}=\sum_{l=1}^{L}p_{l}\psi\left(q_{l}\right)R$
and we must have $\sum_{l=1}^{L}p_{l}\psi\left(q_{l}\right)R\leq W_{b}R_{b}$,
i.e., a per node throughput of $R=\frac{W_{b}R_{b}}{\sum_{l=1}^{L}p_{l}\psi\left(q_{l}\right)}$
is achievable.
\end{IEEEproof}

\subsection{Proof of Theorem \ref{thm:LoCoMIMObound}\label{sub:Proof-of-TheoremCoMIMO}}
\begin{IEEEproof}
Without loss of generality, we focus on the layer 1 CoMP transmission.
Consider an achievable scheme where the nodes in $\mathcal{N}_{C}^{1}$
always transmit to all the nodes in $\mathcal{N}_{C}^{2}$ using CoMP
transmission (for a node in $\mathcal{N}_{C}^{2}$ without requesting
a file with CoMP cache mode, the nodes in $\mathcal{N}_{C}^{1}$ simply
transmit dummy packets to this node). At each time slot, the node
clusters are randomly formed such that each cluster contains $N_{c}$
adjacent Tx nodes in $\mathcal{N}_{C}^{1}$ and $N_{c}$ adjacent
Rx nodes in $\mathcal{N}_{C}^{2}$ that collocated in a square area.
Consider the dual network \cite{Liu_10sTSP_Fairness_rate_polit_WF}
of the original network where the nodes in $\mathcal{N}_{C}^{2}$
act as the transmitters and the nodes in $\mathcal{N}_{C}^{1}$ act
as the receivers. We first study the per cluster average sum rate
of the dual network. Then the results can be transferred to the original
network using the network duality in \cite{Liu_10sTSP_Fairness_rate_polit_WF}. 

Consider the following achievable scheme for the dual network. In
each cluster, the nodes in $\mathcal{N}_{C}^{2}$ whose distance from
the cluster boundary is less than a threshold $d_{b}=\Theta\left(\left(N_{c}r_{0}\right)^{\frac{1}{2\left(\alpha-1\right)}}\right)$
is not allowed to transmit for interference control. The other $\overline{N}_{c}=\Theta\left(N_{c}-N_{c}^{\frac{\alpha}{2\left(\alpha-1\right)}}\right)$
nodes in $\mathcal{N}_{C}^{2}$ are scheduled to transmit at a constant
power $P^{'}=\frac{9W_{c}P}{W+7W_{c}}=\Theta\left(P\right)$. Let
us focus on a reference cluster. Let $\mathcal{G}_{0}^{\textrm{Rx}}$
and $\mathcal{G}_{0}^{\textrm{Tx}}$ denote the set of $N_{c}$ Rx
nodes and the set of $\overline{N}_{c}$ scheduled Tx nodes in the
reference cluster, respectively. Let $\mathcal{G}_{0}^{\textrm{I}}$
denote the set of scheduled Tx (interfering) nodes from other clusters.
By treating the inter-cluster interference as noise, we can achieve
the following per cluster average sum rate
\[
C_{u}=W_{c}\textrm{E}\left[\log\left|\mathbf{I}+P^{'}\mathbf{\Omega}^{-1}\mathbf{H}_{c}\mathbf{H}_{c}^{H}\right|\right],
\]
where $\mathbf{H}_{c}=\left[h_{i,j}\right]_{\forall i\in\mathcal{G}_{0}^{\textrm{Rx}},j\in\mathcal{G}_{0}^{\textrm{Tx}}}\in\mathbb{C}^{N_{c}\times\overline{N}_{c}}$
is the channel matrix of the reference cluster, $\mathbf{\Omega}=P^{'}\sum_{n\in\mathcal{G}_{0}^{\textrm{I}}}\mathbf{h}_{I,n}\mathbf{h}_{I,n}^{H}+W_{c}\mathbf{I}$
is the covariance of the inter-cluster interference plus noise at
the Rx nodes, and $\mathbf{h}_{I,n}=\left[h_{i,n}\right]_{\forall i\in\mathcal{G}_{0}^{\textrm{Rx}}}\in\mathbb{C}^{N_{c}\times1}$
is the channel vector from node $n$ to $\mathcal{G}_{0}^{\textrm{Rx}}$.
Noting that $\mathbf{h}_{I,n},n\in\mathcal{G}_{0}^{\textrm{I}}$ has
independent elements and $\mathbf{h}_{I,n}$ is also independent of
$\mathbf{H}_{c}$, we have $\textrm{E}\left[\mathbf{h}_{I,n}\mathbf{h}_{I,n}^{H}|\mathbf{H}_{c}\right]=\textrm{E}\left[\mathbf{h}_{I,n}\mathbf{h}_{I,n}^{H}\right]=\textrm{diag}\left(\left[r_{i,n}^{-\alpha}\right]_{\forall i\in\mathcal{G}_{0}^{\textrm{Rx}}}\right)$.
Moreover, using the fact that the nearest interfering node in $\mathcal{N}_{C}^{2}$
is at least $d_{b}$ away from the Rx nodes in $\mathcal{N}_{C}^{1}$,
it can be shown that $\sum_{n\in\mathcal{G}_{I}}r_{i,n}^{-\alpha}=O\left(d_{b}^{2-\alpha}\right),\forall i\in\mathcal{G}_{0}^{\textrm{Rx}}$.
Hence 
\begin{equation}
\textrm{E}\left[\mathbf{\Omega}|\mathbf{H}_{c}\right]=\left(O\left(Pd_{b}^{2-\alpha}\right)+W_{c}\right)\mathbf{I}.\label{eq:EOmega}
\end{equation}
Then we have

\begin{eqnarray}
C_{u} & \overset{\textrm{a}}{\geq} & W_{c}\textrm{E}\left[\log\left|\mathbf{I}+P^{'}\textrm{E}\left[\mathbf{\Omega}|\mathbf{H}_{c}\right]^{-1}\mathbf{H}_{c}\mathbf{H}_{c}^{H}\right|\right]\nonumber \\
 & \overset{\textrm{b}}{=} & W_{c}\textrm{E}\left[\log\left|\mathbf{I}+\frac{P^{'}}{W_{c}+O\left(Pd_{b}^{2-\alpha}\right)}\mathbf{H}_{c}\mathbf{H}_{c}^{H}\right|\right]\nonumber \\
 & \overset{\textrm{c}}{=} & W_{c}\textrm{E}\left[\log\left|\mathbf{I}+\frac{P^{'}}{W_{c}}\mathbf{H}_{c}^{H}\mathbf{H}_{c}\right|\right]+O\left(PN_{c}^{\frac{\alpha}{2\left(\alpha-1\right)}}\right),\label{eq:CuE}
\end{eqnarray}
where (\ref{eq:CuE}-a) follows from Jensen's inequality, and (\ref{eq:CuE}-b)
follows from (\ref{eq:EOmega}). (\ref{eq:CuE}-c) is true because
\begin{eqnarray}
 &  & \log\left|\mathbf{I}+\frac{P^{'}}{W_{c}+O\left(P^{'}d_{b}^{2-\alpha}\right)}\mathbf{H}_{c}\mathbf{H}_{c}^{H}\right|\nonumber \\
 & = & \sum_{i=1}^{\overline{N}_{c}}\log\left(1+\frac{P^{'}}{W_{c}+O\left(P^{'}d_{b}^{2-\alpha}\right)}\lambda_{i}\right)\nonumber \\
 & \overset{\textrm{a}}{=} & \sum_{i=1}^{\overline{N}_{c}}\log\left(1+\frac{P^{'}}{W_{c}}\lambda_{i}\right)+O\left(P^{'}\overline{N}_{c}d_{b}^{2-\alpha}\right)\nonumber \\
 & \overset{\textrm{b}}{=} & \log\left|\mathbf{I}+\frac{P^{'}\mathbf{H}_{c}^{H}\mathbf{H}_{c}}{W_{c}}\right|+O\left(PN_{c}^{\frac{\alpha}{2\left(\alpha-1\right)}}\right),\label{eq:Eq15}
\end{eqnarray}
where $\lambda_{i}$ is the $i$-th eigenvalue of $\mathbf{H}_{c}^{H}\mathbf{H}_{c}$,
(\ref{eq:Eq15}-a) follows from the first order Taylor expansion:
$\log\left(1+\frac{P^{'}}{W_{c}+x}\lambda_{i}\right)=\log\left(1+\frac{P^{'}}{W_{c}}\lambda_{i}\right)+O\left(x\right)$,
(\ref{eq:Eq15}-b) follows from $\log\left|\mathbf{I}+\frac{P^{'}\mathbf{H}_{c}^{H}\mathbf{H}_{c}}{W_{c}}\right|=\sum_{i=1}^{\overline{N}_{c}}\log\left(1+\frac{P^{'}}{W_{c}}\lambda_{i}\right)$,
$P^{'}=\Theta\left(P\right)$, $\overline{N}_{c}=O\left(N_{c}\right)$
and $d_{b}=\Theta\left(N_{c}^{\frac{1}{2\left(\alpha-1\right)}}\right)$.

We can use the same technique as in Appendix I of \cite{Tse_IT07_CapscalingHMIMO}
to bound the term $\overline{C}_{u}\triangleq W_{c}\textrm{E}\left[\log\left|\mathbf{I}+\frac{P^{'}}{W_{c}}\mathbf{H}_{c}^{H}\mathbf{H}_{c}\right|\right]$.
Let $\lambda$ be chosen uniformly among the $\overline{N}_{c}$ eigenvalues
of $\frac{\mathbf{H}_{c}^{H}\mathbf{H}_{c}}{\overline{N}_{c}}$. Then
\begin{eqnarray}
\overline{C}_{u} & \geq & W_{c}\overline{N}_{c}\textrm{E}\left[\log\left(1+\frac{\overline{N}_{c}P^{'}}{W_{c}}\lambda\right)\right]\nonumber \\
 & \geq & W_{c}\overline{N}_{c}\log\left(1+\frac{\overline{N}_{c}P^{'}}{W_{c}}t\right)\Pr\left(\lambda>t\right),\label{eq:Cubar}
\end{eqnarray}
for any $t\geq0$. By the Paley-Zygmund inequality, we have 
\begin{equation}
\Pr\left(\lambda>t\right)\geq\frac{\left(\textrm{E}\left(\lambda\right)-t\right)^{2}}{\textrm{E}\left(\lambda^{2}\right)},\:0\leq t<\textrm{E}\left(\lambda\right).\label{eq:Prlamt}
\end{equation}
Let $\overline{\mathcal{G}}_{0}^{\textrm{Rx}}$ denote the set of
all Rx nodes excluding the Rx nodes in $\mathcal{G}_{0}^{\textrm{Rx}}$.
We have 
\begin{eqnarray}
\sum_{i\in\overline{\mathcal{G}}_{0}^{\textrm{Rx}}}r_{i,k}^{-\alpha} & \underset{\leq}{\textrm{a}} & 4r_{0}^{-\alpha}\sum_{j=\left\lceil \frac{d_{b}}{r_{0}}\right\rceil }^{\infty}\sum_{i=-\infty}^{\infty}\left(i^{2}+j^{2}\right)^{-\frac{\alpha}{2}}\nonumber \\
 & = & \Theta\left(\int_{\left\lceil \frac{d_{b}}{r_{0}}\right\rceil }^{\infty}\int_{-\infty}^{\infty}\left(x^{2}+y^{2}\right)^{-\frac{\alpha}{2}}dxdy\right)\nonumber \\
 & \overset{\textrm{b}}{=} & \Theta\left(\frac{2\sqrt{\pi}\mathbb{G}\left(\frac{\alpha-1}{2}\right)\left\lceil \frac{d_{b}}{r_{0}}\right\rceil ^{2-\alpha}}{(\alpha-2)^{2}\mathbb{G}\left(\frac{\alpha}{2}-1\right)}\right)\nonumber \\
 & = & \Theta\left(N_{c}^{-\frac{\alpha-2}{2\left(\alpha-1\right)}}\right)\overset{\textrm{c}}{=}\Theta\left(d_{b}^{2-\alpha}\right),\forall k\in\mathcal{G}_{0}^{\textrm{Tx}},\label{eq:Eq18}
\end{eqnarray}
where (\ref{eq:Eq18}-a) is true because the nearest node in $\overline{\mathcal{G}}_{0}^{\textrm{Rx}}$
is at least $d_{b}$ away from a node in $\mathcal{G}_{0}^{\textrm{Tx}}$,
$\mathbb{G}\left(\cdot\right)$ is the Gamma function and (\ref{eq:Eq18}-b)
follows from a direct calculation of the two-dimensional integration,
(\ref{eq:Eq18}-c) follows from $d_{b}=\Theta\left(\left(N_{c}r_{0}\right)^{\frac{1}{2\left(\alpha-1\right)}}\right)$.
It follows from (\ref{eq:Eq18}) that 
\begin{equation}
\sum_{i\in\overline{\mathcal{G}}_{0}^{\textrm{Rx}}}r_{i,k}^{-\alpha}=O\left(N_{c}^{-\frac{\alpha-2}{2\left(\alpha-1\right)}}\right),\forall k\in\mathcal{G}_{0}^{\textrm{Tx}}.\label{eq:sumG0barRx}
\end{equation}
Then it follows from (\ref{eq:sumG0barRx}) and $\sum_{i\in\mathcal{G}_{0}^{\textrm{Rx}}}r_{i,k}^{-\alpha}=G_{C}-\sum_{i\in\overline{\mathcal{G}}_{0}^{\textrm{Rx}}}r_{i,k}^{-\alpha}$
that
\begin{equation}
\sum_{i\in\mathcal{G}_{0}^{\textrm{Rx}}}r_{i,k}^{-\alpha}=G_{C}+O\left(N_{c}^{-\frac{\alpha-2}{2\left(\alpha-1\right)}}\right),\forall k\in\mathcal{G}_{0}^{\textrm{Tx}},\label{eq:sumGrx}
\end{equation}
where $G_{C}\triangleq\sum_{i\in\mathcal{G}_{0}^{\textrm{Rx}}\cup\overline{\mathcal{G}}_{0}^{\textrm{Rx}}}r_{i,k}^{-\alpha}$
is defined in Theorem \ref{thm:LoCoMIMObound}. Following similar
analysis as in Appendix I of \cite{Tse_IT07_CapscalingHMIMO}, we
have 
\begin{eqnarray}
\textrm{E}\left(\lambda\right) & = & \frac{1}{\overline{N}_{c}}\textrm{E}\left(\textrm{Tr}\left(\frac{\mathbf{H}_{c}^{H}\mathbf{H}_{c}}{\overline{N}_{c}}\right)\right)\nonumber \\
 & = & \frac{1}{\overline{N}_{c}^{2}}\sum_{k\in\mathcal{G}_{0}^{\textrm{Tx}}}\sum_{i\in\mathcal{G}_{0}^{\textrm{Rx}}}\textrm{E}\left(\left|h_{i,k}\right|^{2}\right)\nonumber \\
 & = & \frac{1}{\overline{N}_{c}^{2}}\sum_{k\in\mathcal{G}_{0}^{\textrm{Tx}}}\sum_{i\in\mathcal{G}_{0}^{\textrm{Rx}}}r_{i,k}^{-\alpha}\nonumber \\
 & = & \frac{G_{C}+O\left(N_{c}^{-\frac{\alpha-2}{2\left(\alpha-1\right)}}\right)}{\overline{N}_{c}},\label{eq:Elam}
\end{eqnarray}
\begin{eqnarray}
\textrm{E}\left(\lambda^{2}\right) & = & \frac{1}{\overline{N}_{c}}\textrm{E}\left(\textrm{Tr}\left(\frac{\mathbf{H}_{c}^{H}\mathbf{H}_{c}\mathbf{H}_{c}^{H}\mathbf{H}_{c}}{\overline{N}_{c}^{2}}\right)\right)\nonumber \\
 & = & \frac{1}{\overline{N}_{c}^{3}}\sum_{k\in\mathcal{G}_{0}^{\textrm{Tx}}}\sum_{i\in\mathcal{G}_{0}^{\textrm{Rx}}}\sum_{m\in\mathcal{G}_{0}^{\textrm{Tx}}}\sum_{l\in\mathcal{G}_{0}^{\textrm{Rx}}}\textrm{E}\left(h_{i,k}h_{l,k}^{*}h_{l,m}h_{i,m}^{*}\right)\nonumber \\
 & \leq & \frac{2}{\overline{N}_{c}^{3}}\sum_{k\in\mathcal{G}_{0}^{\textrm{Tx}}}\sum_{i\in\mathcal{G}_{0}^{\textrm{Rx}}}\sum_{l\in\mathcal{G}_{0}^{\textrm{Rx}}}\textrm{E}\left(\left|h_{i,k}\right|^{2}\right)\textrm{E}\left(\left|h_{l,k}\right|^{2}\right)\nonumber \\
 & = & \frac{2}{\overline{N}_{c}^{3}}\sum_{k\in\mathcal{G}_{0}^{\textrm{Tx}}}\sum_{i\in\mathcal{G}_{0}^{\textrm{Rx}}}r_{i,k}^{-\alpha}\sum_{l\in\mathcal{G}_{0}^{\textrm{Rx}}}r_{l,k}^{-\alpha}\nonumber \\
 & = & \frac{2\left(G_{C}+O\left(N_{c}^{-\frac{\alpha-2}{2\left(\alpha-1\right)}}\right)\right)^{2}}{\overline{N}_{c}^{2}}.\label{eq:varlam}
\end{eqnarray}
where the last equalities in the above two equations follow from (\ref{eq:sumGrx})
and $\sum_{k\in\mathcal{G}_{0}^{\textrm{Tx}}}=\overline{N}_{c}$.
Choosing $t=\frac{r_{0}^{-\alpha}}{\overline{N}_{c}}$ and using (\ref{eq:Cubar}),
(\ref{eq:Prlamt}), (\ref{eq:Elam}) and (\ref{eq:varlam}), we have
\[
\overline{C}_{u}\geq W_{c}\overline{N}_{c}\rho\log\left(1+\frac{P^{'}r_{0}^{-\alpha}}{W_{c}}\right)+O\left(PN_{c}^{\frac{\alpha}{2\left(\alpha-1\right)}}\right).
\]

According to the network duality in \cite{Liu_10sTSP_Fairness_rate_polit_WF},
a per cluster average sum rate of $C_{d}=C_{u}\geq\overline{C}_{u}+O\left(PN_{c}^{\frac{\alpha}{2\left(\alpha-1\right)}}\right)$
can be achieved in the original network with equal or less total network
power. Since the clusters are randomly formed, all nodes in $\mathcal{N}_{C}^{1}$
(or in $\mathcal{N}_{C}^{2}$) are statistically symmetric. As a result,
the per node average rate $R_{c}^{'}$ that can be provided by the
PHY is lower bounded as 

\begin{eqnarray}
R_{c}^{'} & \geq & \frac{C_{d}}{N_{c}}\geq\rho W_{c}\log\left(1+\frac{P^{'}r_{0}^{-\alpha}}{W_{c}}\right)+O\left(PN_{c}^{-\frac{\alpha-2}{2\left(\alpha-1\right)}}\right)\nonumber \\
 & \overset{\textrm{a}}{\geq} & \rho W_{c}\log\left(1+\frac{2Pr_{0}^{-\alpha}}{W}\right)+O\left(PN_{c}^{-\frac{\alpha-2}{2\left(\alpha-1\right)}}\right)\label{eq:RcpL}
\end{eqnarray}
and the average power at each node in $\mathcal{N}_{C}^{1}$ required
to achieve the above per node average rate is no more than $P^{'}$,
where (\ref{eq:RcpL}-a) follows from $\frac{P^{'}}{W_{c}}=\frac{9P}{W+7W_{c}}\geq\frac{2P}{W}$.
On the other hand, using the cut set bound between all nodes in $\mathcal{N}_{C}^{1}$
and a node in $\mathcal{N}_{C}^{2}$, we have
\begin{equation}
R_{c}^{'}\leq W_{c}\log\left(1+\frac{P^{'}G_{C}}{W_{c}}\right)\overset{\textrm{a}}{\leq}W_{c}\log\left(1+\frac{9PG_{C}}{W}\right),\label{eq:RcpU}
\end{equation}
where (\ref{eq:RcpU}-a) follows from $\frac{P^{'}}{W_{c}}=\frac{9P}{W+7W_{c}}\leq\frac{9P}{W}$.
\end{IEEEproof}

\subsection{Proof of Corollary \ref{cor:PBA}\label{sub:Proof-of-CorollaryPBA}}
\begin{IEEEproof}
Let us first consider the case when $Q_{\Omega\left(\mathbf{q}\right)}>0$
and $Q_{\overline{\Omega}\left(\mathbf{q}\right)}>0$. Consider the
transmission of $L_{0}$ files to a reference node $n_{0}$. Following
similar analysis as that in Appendix \ref{sub:Proof-of-TheoremPBTA},
it can be shown that the\textit{ per link multihop traffic} (i.e.,
the traffic of each link on the multihop band) induced by transmitting
files with multihop cache mode is $\sum_{l\in\Omega\left(\mathbf{q}\right)}FL_{l}\psi\left(q_{l}\right)$,
where $L_{l}$ is the number of transmitting the $l$-th file. Clearly,
the \textit{CoMP traffic} (i.e., the traffic delivered to node $n_{0}$
on CoMP band) induced by transmitting files with CoMP cache mode is
$\sum_{l\in\overline{\Omega}\left(\mathbf{q}\right)}FL_{l}\left(1-q_{l}\right)$.
The transmission of $L_{0}$ files can be finished within time $t_{0}$
if and only if both the per link multihop traffic can be delivered
on the multihop band and the CoMP traffic can be delivered on the
CoMP band, i.e., $\sum_{l\in\Omega\left(\mathbf{q}\right)}FL_{l}\psi\left(q_{l}\right)\leq\left(W-2W_{c}\right)R_{m}\left(W_{c}\right)t_{0}$
and $\sum_{l\in\overline{\Omega}\left(\mathbf{q}\right)}FL_{l}\left(1-q_{l}\right)\leq W_{c}R_{c}\left(W_{c}\right)t_{0}$.
Hence, the per node throughput for fixed bandwidth partition $W_{c},W_{b}=W-2W_{c}$
is
\begin{eqnarray*}
 &  & T_{\mathbf{q}}\left(W_{c}\right)\\
 & = & \lim_{L_{0}\rightarrow\infty}\frac{L_{0}F}{t_{0}}\\
 & = & \lim_{L_{0}\rightarrow\infty}\frac{L_{0}F}{\max\left(\frac{\sum_{l\in\Omega\left(\mathbf{q}\right)}FL_{l}\psi\left(q_{l}\right)}{\left(W-2W_{c}\right)R_{m}\left(W_{c}\right)},\frac{\sum_{l\in\overline{\Omega}\left(\mathbf{q}\right)}FL_{l}\left(1-q_{l}\right)}{W_{c}R_{c}\left(W_{c}\right)}\right)}\\
 & = & \lim_{L_{0}\rightarrow\infty}\frac{1}{\max\left(\frac{Q_{\Omega\left(\mathbf{q}\right)}}{\left(W-2W_{c}\right)R_{m}\left(W_{c}\right)},\frac{Q_{\overline{\Omega}\left(\mathbf{q}\right)}}{W_{c}R_{c}\left(W_{c}\right)}\right)},
\end{eqnarray*}
where the last equality follows from $\lim_{L_{0}\rightarrow\infty}\frac{L_{l}}{L_{0}}=p_{l}$.
Hence, the per node throughput $\Gamma_{A}\left(\mathbf{q}\right)=\max_{W_{c}\in\left(0,W_{c}\right)}T_{\mathbf{q}}\left(W_{c}\right)=T_{\mathbf{q}}\left(W_{c}^{*}\right)$,
where the optimal solution $W_{c}^{*}$ is the unique solution of
(\ref{eq:Wstar}). Using (\ref{eq:Wstar}), it can be verified that
$T_{\mathbf{q}}\left(W_{c}^{*}\right)$ is given by (\ref{eq:TAqgen}). 

It can be verified that $\Gamma_{A}\left(\mathbf{q}\right)$ is still
given by (\ref{eq:Taqgen}) when $Q_{\Omega\left(\mathbf{q}\right)}=0$
or $Q_{\overline{\Omega}\left(\mathbf{q}\right)}=0$. Moreover, $\Gamma_{A}^{U}\left(\mathbf{q}\right)\geq\Gamma_{A}\left(\mathbf{q}\right)\geq\Gamma_{A}^{L}\left(\mathbf{q}\right)+O\left(PN_{c}^{-\frac{\alpha-2}{2\left(\alpha-1\right)}}\right)$
follows from that $\Gamma_{A}\left(\mathbf{q}\right)$ is an increasing
function of $R_{m}\left(W_{c}^{*}\right)$ and $R_{c}\left(W_{c}^{*}\right)$.
Finally, as $P,N_{c}\rightarrow\infty$ such that $PN_{c}^{-\frac{\alpha-2}{2\left(\alpha-1\right)}}\rightarrow0$,
we have $R_{m}^{a}\rightarrow\frac{1}{9}\log\left(1+\frac{r_{0}^{-\alpha}}{I_{R}}\right)$
and $R_{m}^{a}/R_{c}^{a}\rightarrow0,$ for $a\in\left\{ L,U\right\} $,
from which (\ref{eq:TAq}) follows.
\end{IEEEproof}

\subsection{Proof of Corollary \ref{cor:OthpR} \label{sub:Proof-of-CorollaryOthpR}}
\begin{IEEEproof}
When $q_{l}=o\left(1\right)$, it can be shown that $\psi\left(q_{l}\right)=\Theta\left(\sqrt{\frac{1}{q_{l}}}\right)$.
When $q_{l}=\Theta\left(1\right)$, it can be shown that $\psi\left(q_{l}\right)=O\left(1\right)$.
Let $\mathcal{L}_{1}=\left\{ l:\: q_{l}=\Theta\left(1\right)\right\} $.
Since $\sum_{l=1}^{L}q_{l}\leq\frac{B_{C}}{F}$, the cardinality of
$\mathcal{L}_{1}$ must be bounded, i.e., $\left|\mathcal{L}_{1}\right|=\Theta\left(1\right)$.
As a result, we have $\sum_{l\in\mathcal{L}_{1}}p_{l}\psi\left(q_{l}\right)=O\left(1\right)$
and 
\begin{eqnarray}
 &  & \sum_{l=1}^{L}p_{l}\psi\left(q_{l}\right)\nonumber \\
 & = & \Theta\left(\sum_{l\notin\mathcal{L}_{1}}p_{l}\sqrt{\frac{1}{q_{l}}}\right)+\sum_{l\in\mathcal{L}_{1}}p_{l}\psi\left(q_{l}\right)\nonumber \\
 & = & \Theta\left(\sum_{l}p_{l}\sqrt{\frac{1}{q_{l}}}\right)-\Theta\left(\sum_{l\in\mathcal{L}_{1}}p_{l}\sqrt{\frac{1}{q_{l}}}\right)+O\left(1\right)\nonumber \\
 & = & \Theta\left(\sum_{l}p_{l}\sqrt{\frac{1}{q_{l}}}\right)+O\left(1\right)=\Theta\left(\sum_{l}p_{l}\sqrt{\frac{1}{q_{l}}}\right),\label{eq:sumplfai}
\end{eqnarray}
where the last equality holds because $\Theta\left(\sum_{l}p_{l}\sqrt{\frac{1}{q_{l}}}\right)\geq\Theta\left(1\right)$.
It follows from (\ref{eq:sumplfai}) and Theorem \ref{thm:PBTA} that
$\Gamma_{B}\left(\mathbf{q}\right)=\Theta\left(\frac{1}{\sum_{l=1}^{L}p_{l}\sqrt{\frac{1}{q_{l}}}}\right)$.
For bounded power $P$ and $\alpha>2$, it can be shown that the interference
seen at any node is bounded, from which it follows that $R_{c}\left(W_{c}\right)=\Theta\left(1\right)$.
Hence, $\Gamma_{A}\left(\mathbf{q}\right)=\Theta\left(\frac{1}{Q_{\Omega\left(\mathbf{q}\right)}+2Q_{\overline{\Omega}\left(\mathbf{q}\right)}}\right)=\Theta\left(\sum_{l}p_{l}\sqrt{\frac{1}{q_{l}}}\right)$,
where the last equality follows similar analysis as in (\ref{eq:sumplfai}).
\end{IEEEproof}

\subsection{Proof of Theorem \ref{thm:Order-optimal-pR} \label{sub:Proof-of-Theorem-OdpR}}
\begin{IEEEproof}
Consider the following relaxed problem:
\begin{equation}
\min_{\mathbf{q}>0}\: f\left(\mathbf{q}\right)\triangleq\sum_{l=1}^{L}p_{l}\sqrt{\frac{1}{q_{l}}},\: s.t.\:\sum_{l=1}^{L}q_{l}\leq\frac{B_{C}}{F}.\label{eq:optq-1}
\end{equation}
Using the Lagrange dual method, it can be shown that the optimal solution
and the optimal value of Problem (\ref{eq:optq-1}) are given by $q_{l}^{\star}=\frac{B_{C}p_{l}^{2/3}}{F\sum_{l=1}^{L}p_{l}^{2/3}},\forall l$
and $f\left(\mathbf{q}^{\star}\right)=\sqrt{\frac{F}{B_{C}}}\left[\sum_{l=1}^{L}p_{l}^{2/3}\right]^{3/2}$,
respectively. On the other hand, the objective value of Problem (\ref{eq:optq})
with cache content replication vector $\mathbf{q}^{*}$ in (\ref{eq:odroptq})
is given by $f\left(\mathbf{q}^{*}\right)\leq f\left(\mathbf{q}^{*}-\frac{1}{N}\right)=\Theta\left(\left[\sum_{l=1}^{L}p_{l}^{2/3}\right]^{3/2}\right)$.
Let $f^{*}$ denote the optimal value of Problem (\ref{eq:optq}).
Since $f\left(\mathbf{q}^{\star}\right)\leq f^{*}\leq f\left(\mathbf{q}^{*}\right)$,
$f\left(\mathbf{q}^{\star}\right)=\Theta\left(\left[\sum_{l=1}^{L}p_{l}^{2/3}\right]^{3/2}\right)$
and $f\left(\mathbf{q}^{*}\right)\leq\Theta\left(\left[\sum_{l=1}^{L}p_{l}^{2/3}\right]^{3/2}\right)$,
we have $f^{*}=\Theta\left(\left[\sum_{l=1}^{L}p_{l}^{2/3}\right]^{3/2}\right)$,
which proves that $\mathbf{q}^{*}$ is the order-optimal cache content
replication vector. 
\end{IEEEproof}

\subsection{Proof of Theorem \ref{thm:CLB}\label{sub:Proof-of-TheoremCLB}}

Consider a simple bandwidth partition: $W_{b}=W_{c}=\frac{W}{3}$.
We first prove that a throughput of $\Theta\left(\frac{C_{B}}{\sum_{l=1}^{L}p_{l}\sqrt{\frac{1}{q_{l}}}}\right)$,
where $C_{B}\triangleq\frac{W}{M}\log\left(1+\frac{3MP\left(2r_{\textrm{max}}\right)^{-\alpha}}{\left(M+1\right)W+3MI_{A}}\right)$,
is achievable for node $n$ requesting a file $l_{n}$ with multihop
cache mode. The proof relies on Lemma \ref{lem:FRboundM} and the
following lemma.
\begin{lem}
[Upper bound of source radius]\label{lem:SBSbound}For a node $n$
requesting a file $l_{n}$ with multihop cache mode 
\[
r_{n}^{*}\leq\left(2\sqrt{\left\lceil 1/q_{l_{n}}\right\rceil -1}+1\right)r_{\textrm{max}}.
\]
\end{lem}
\begin{IEEEproof}
Let $\mathcal{O}_{n}$ denote a disk centered at the node $n$ with
radius $r_{n}^{*}-r_{\textrm{max}}$ and let $\mathcal{O}_{n}^{'}$
denote the intersection of $\mathcal{O}_{n}$ and the network coverage
area (i.e., the square of area $Nr_{0}^{2}$). By Assumption \ref{asm:Place}-2),
any point inside $\mathcal{O}_{n}^{'}$ must lie in the Voronoi cells
corresponding to the nodes in $\overline{\mathcal{B}}_{n}\cup n$.
Since the area of each Voronoi cell is less than $\pi r_{\textrm{max}}^{2}$,
we must have $\mathcal{A}\left(\mathcal{O}_{n}^{'}\right)\leq\left(\left|\overline{\mathcal{B}}_{n}\right|+1\right)\pi r_{\textrm{max}}^{2}$,
where $\mathcal{A}\left(\mathcal{O}_{n}^{'}\right)$ denotes the area
of $\mathcal{O}_{n}^{'}$. Since $\mathcal{A}\left(\mathcal{O}_{n}^{'}\right)\geq\frac{\pi\left(r_{n}^{*}-r_{\textrm{max}}\right)^{2}}{4}$
and $\left|\overline{\mathcal{B}}_{n}\right|\leq\left\lceil 1/q\right\rceil -2$,
we have $\frac{\pi\left(r_{n}^{*}-r_{\textrm{max}}\right)^{2}}{4}\leq\left(\left\lceil 1/q\right\rceil -1\right)\pi r_{\textrm{max}}^{2}$
and thus $r_{n}^{*}\leq\left(2\sqrt{\left\lceil 1/q\right\rceil -1}+1\right)r_{\textrm{max}}$.
\end{IEEEproof}

We first derive a lower bound for the throughput of each link on the
multihop band. For each link, the bandwidth is $\frac{W}{3M}$ and
the transmit power on this bandwidth is $\frac{P}{1+M}$. The noise
power is $\frac{W}{3M}$ and the interference power is upper bounded
by $\frac{P}{1+M}I_{A}$ as will be shown later. The channel gain
of each link is lower bounded by $\left(2r_{\textrm{max}}\right)^{-\alpha}$
since the distance between any two adjacent nodes is upper bounded
by $2r_{\textrm{max}}$ by Assumption \ref{asm:Place}-2). It follows
that the throughput of each link is lower bounded by $\frac{C_{B}}{3}$. 

Suppose we want to support a per node throughput of $R$. Using Assumption
\ref{asm:Place} and Lemma \ref{lem:SBSbound}, it can be shown that
the average traffic $T_{n}$ to be relayed by the $n$-th node on
the multihop band is upper bounded as $T_{n}\leq\Theta\left(\sum_{l=1}^{L}p_{l}\sqrt{\frac{1}{q_{l}}}R\right)$
for all $n$. Clearly, a per node throughput of $R$ can be supported
if no node is overloaded, i.e., $T_{n}\leq\frac{C_{B}}{3},\forall n$.
Hence, a per node throughput of $R=\Theta\left(\frac{C_{B}}{\sum_{l=1}^{L}p_{l}\sqrt{\frac{1}{q_{l}}}}\right)$
is achievable.

The rest is to prove that the interference power $I_{B}$ seen at
each node on the multihop band is upper bounded by $\frac{1}{3M}I_{A}$.
For any node $n$, let $\mathcal{O}_{n}\left(ix\right)$ denote a
disk centered at the node $n$ with radius $ix$ and define a set
of rings $\mathcal{O}_{n,i}\triangleq\mathcal{O}_{n}\left(\left(i+1\right)x\right)\backslash\mathcal{O}_{n}\left(ix\right),i=1,2,...$.
Using Assumption \ref{asm:Place}-1), it can be shown that the number
of nodes in the $i$-th ring $\mathcal{O}_{n,i}$ is upper bounded
by $\frac{\left(\left(i+1\right)x+\frac{r_{\textrm{min}}}{2}\right)^{2}-\left(ix-\frac{r_{\textrm{min}}}{2}\right)^{2}}{r_{\textrm{min}}^{2}/4}$.
Clearly, the distance between a node in $\mathcal{O}_{n,i}$ and node
$n$ is lower bounded by $ix$. Moreover, from Assumption \ref{asm:Place}-2)
and Condition \ref{cond:FR}, there is no interference from the nodes
in $\mathcal{O}_{n}\left(r_{I}-2r_{\textrm{max}}\right)$ to node
$n$. As a result, if we let $x=r_{I}-2r_{\textrm{max}}$, the interference
power $I_{B}$ seen by any node can be upper bounded as
\begin{eqnarray*}
I_{B} & \leq & \sum_{i=1}^{\infty}\frac{\left(\left(i+1\right)x+\frac{r_{\textrm{min}}}{2}\right)^{2}-\left(ix-\frac{r_{\textrm{min}}}{2}\right)^{2}}{r_{\textrm{min}}^{2}/4}\frac{P}{\left(1+M\right)\left(ix\right)^{\alpha}}\\
 & \leq & \frac{4P\left(x+r_{\textrm{min}}\right)}{\left(1+M\right)r_{\textrm{min}}^{2}}\sum_{i=1}^{\infty}\left(\frac{2}{\left(ix\right)^{\alpha-1}}+\frac{1}{i^{\alpha}x^{\alpha-1}}\right)\\
 & \leq & \frac{4P\left(x+r_{\textrm{min}}\right)}{\left(1+M\right)r_{\textrm{min}}^{2}}\left[\int_{1}^{\infty}\left(\frac{2}{\left(zx\right)^{\alpha-1}}+\frac{1}{z^{\alpha}x^{\alpha-1}}\right)dz+\frac{3}{x^{\alpha-1}}\right]\\
 & = & \frac{4P\left(x+r_{\textrm{min}}\right)}{\left(1+M\right)r_{\textrm{min}}^{2}x^{\alpha-1}}\left(\frac{2}{\alpha-2}+\frac{1}{\alpha-1}+3\right).
\end{eqnarray*}

On the other hand, when node $n$ requests a file $l_{n}$ with CoMP
cache mode, it is clear that a throughput of $\Theta\left(1\right)\geq\Theta\left(\frac{C_{B}}{\sum_{l=1}^{L}p_{l}\sqrt{\frac{1}{q_{l}}}}\right)$
is achievable for node $n$. As a result, an overall throughput of
$R=\Theta\left(\frac{C_{B}}{\sum_{l=1}^{L}p_{l}\sqrt{\frac{1}{q_{l}}}}\right)$
is also achievable for node $n$.



\begin{thebibliography}{10}
\providecommand{\url}[1]{#1}
\csname url@samestyle\endcsname
\providecommand{\newblock}{\relax}
\providecommand{\bibinfo}[2]{#2}
\providecommand{\BIBentrySTDinterwordspacing}{\spaceskip=0pt\relax}
\providecommand{\BIBentryALTinterwordstretchfactor}{4}
\providecommand{\BIBentryALTinterwordspacing}{\spaceskip=\fontdimen2\font plus
\BIBentryALTinterwordstretchfactor\fontdimen3\font minus
  \fontdimen4\font\relax}
\providecommand{\BIBforeignlanguage}[2]{{%
\expandafter\ifx\csname l@#1\endcsname\relax
\typeout{** WARNING: IEEEtran.bst: No hyphenation pattern has been}%
\typeout{** loaded for the language `#1'. Using the pattern for}%
\typeout{** the default language instead.}%
\else
\language=\csname l@#1\endcsname
\fi
#2}}
\providecommand{\BIBdecl}{\relax}
\BIBdecl

\bibitem{Goebbels_IJCS10_Smartcaching}
S.~Goebbels, ``Disruption tolerant networking by smart caching,'' \emph{Int. J.
  Commun. Syst.}, vol.~23, no.~5, pp. 569--595, May 2010.

\bibitem{Caire_INFOCOM12_femtocache}
N.~Golrezaei, K.~Shanmugam, A.~Dimakis, A.~Molisch, and G.~Caire,
  ``Femtocaching: Wireless video content delivery through distributed caching
  helpers,'' \emph{Proc. IEEE INFOCOM}, pp. 1107--1115, 2012.

\bibitem{Kozat_TOM09_P2P}
U.~Kozat, O.~Harmanci, S.~Kanumuri, M.~Demircin, and M.~Civanlar, ``Peer
  assisted video streaming with supply-demand-based cache optimization,''
  \emph{IEEE Transactions on Multimedia}, vol.~11, no.~3, pp. 494--508, 2009.

\bibitem{Shen_TOM04_CDN}
B.~Shen, S.-J. Lee, and S.~Basu, ``Caching strategies in transcoding-enabled
  proxy systems for streaming media distribution networks,'' \emph{IEEE
  Transactions on Multimedia}, vol.~6, no.~2, pp. 375--386, 2004.

\bibitem{Xiaohu_TWC11_COMIMO}
X.~Ge, K.~Huang, C.-X. Wang, X.~Hong, and X.~Yang, ``Capacity analysis of a
  multi-cell multi-antenna cooperative cellular network with co-channel
  interference,'' \emph{IEEE Trans. Wireless Commun.}, vol.~10, no.~10, pp.
  3298--3309, October 2011.

\bibitem{Niesen_TIT12_FLcaching}
M.~Maddah-Ali and U.~Niesen, ``Fundamental limits of caching,'' \emph{IEEE
  Trans. Info. Theory}, vol.~60, no.~5, pp. 2856--2867, May 2014.

\bibitem{Debbah_CoM14_proactivecache}
E.~Bastug, M.~Bennis, and M.~Debbah, ``Living on the edge: The role of
  proactive caching in {5G} wireless networks,'' \emph{IEEE Communications
  Magazine}, vol.~52, no.~8, pp. 82--89, Aug 2014.

\bibitem{Gitzenis_TIT13_wirelesscache}
S.~Gitzenis, G.~Paschos, and L.~Tassiulas, ``Asymptotic laws for joint content
  replication and delivery in wireless networks,'' \emph{IEEE Trans. Info.
  Theory}, vol.~59, no.~5, pp. 2760--2776, May 2013.

\bibitem{Caire_arxiv13_D2Dcaching}
\BIBentryALTinterwordspacing
M.~Ji, G.~Caire, and A.~Molisch, ``Fundamental limits of distributed caching in
  {D2D} wireless networks,'' 2013. [Online]. Available:
  \url{http://arxiv.org/abs/1304.5856}
\BIBentrySTDinterwordspacing

\bibitem{Caire_arxiv13_d2dcachingtradeoff}
\BIBentryALTinterwordspacing
M.~Ji, G.~Caire, and A.~F. Molisch, ``The throughput-outage tradeoff of
  wireless one-hop caching networks,'' 2013. [Online]. Available:
  \url{http://arxiv.org/abs/1312.2637}
\BIBentrySTDinterwordspacing

\bibitem{Caire_arxiv13_d2dcachingtutorial}
\BIBentryALTinterwordspacing
------, ``Wireless device-to-device caching networks: Basic principles and
  system performance,'' 2013. [Online]. Available:
  \url{http://arxiv.org/abs/1305.5216}
\BIBentrySTDinterwordspacing

\bibitem{Altieri_arxiv14_d2dcaching}
\BIBentryALTinterwordspacing
A.~Altieri, P.~Piantanida, L.~R. Vega, and C.~Galarza, ``On fundamental
  trade-offs of device-to-device communications in large wireless networks,''
  2014. [Online]. Available: \url{http://arxiv.org/abs/1405.2295}
\BIBentrySTDinterwordspacing

\bibitem{Jeon_ICC15_D2Dcaching}
S.-W. Jeon, S.-N. Hong, M.~Ji, and G.~Caire, ``Caching in wireless multihop
  device-to-device networks,'' in \emph{to appear in IEEE ICC 2015}, Jun 2015.

\bibitem{GuptaKumar}
P.~Gupta and P.~Kumar, ``The capacity of wireless networks,'' \emph{IEEE Trans.
  Info. Theory}, vol.~46, no.~2, pp. 388--404, Mar 2000.

\bibitem{Liu_TSP14_CacheRelay}
A.~Liu and V.~Lau, ``Cache-enabled opportunistic cooperative {MIMO} for video
  streaming in wireless systems,'' \emph{IEEE Trans. Signal Processing},
  vol.~62, no.~2, pp. 390--402, Jan 2014.

\bibitem{Liu_TSP13_CacheIFN}
------, ``Mixed-timescale precoding and cache control in cached {MIMO}
  interference network,'' \emph{IEEE Trans. Signal Processing}, vol.~61,
  no.~24, pp. 6320--6332, Dec 2013.

\bibitem{Breslau_INFOCOM99_ZipfLaw}
L.~Breslau, P.~Cao, L.~Fan, G.~Phillips, and S.~Shenker, ``Web caching and
  zipf-like distributions: evidence and implications,'' in \emph{Proc. INFOCOM,
  New York, USA}, Mar 1999, pp. 126--134.

\bibitem{Yamakami_PDCAT06_Zipflaw}
T.~Yamakami, ``A zipf-like distribution of popularity and hits in the mobile
  web pages with short life time,'' in \emph{Proc. Parallel Distrib. Comput.,
  Appl. Technol., Taipei, Taiwan}, Dec 2006, pp. 240--243.

\bibitem{Tse_IT07_CapscalingHMIMO}
A.~Ozgur, O.~Leveque, and D.~Tse, ``Hierarchical cooperation achieves optimal
  capacity scaling in ad hoc networks,'' \emph{IEEE Trans. Info. Theory},
  vol.~53, no.~10, pp. 3549--3572, Oct 2007.

\bibitem{Niesen_TIT10_CSadhoc}
U.~Niesen, P.~Gupta, and D.~Shah, ``The balanced unicast and multicast capacity
  regions of large wireless networks,'' \emph{IEEE Trans. Info. Theory},
  vol.~56, no.~5, pp. 2249--2271, May 2010.

\bibitem{Iacobelli_CIP10_DFA}
L.~Iacobelli, F.~Scoubart, D.~Pirez, P.~Fouillot, R.~Massin, C.~Lefebvre,
  C.~Le~Martret, and V.~Conan, ``Dynamic frequency allocation in ad hoc
  networks,'' in \emph{2010 2nd International Workshop on Cognitive Information
  Processing (CIP)}, June 2010, pp. 145--150.

\bibitem{Elsner_ICUMT10_DFA}
J.~Elsner, R.~Tanbourgi, and F.~Jondral, ``On the transmission capacity of
  wireless multi-channel ad hoc networks with local {FDMA} scheduling,'' in
  \emph{2010 International Congress on Ultra Modern Telecommunications and
  Control Systems and Workshops (ICUMT)}, Oct 2010, pp. 315--320.

\bibitem{weingarten2006capacity}
H.~Weingarten, Y.~Steinberg, and S.~Shamai, ``The capacity region of the
  gaussian multiple-input multiple-output broadcast channel,'' \emph{IEEE
  Trans. Info. Theory}, vol.~52, no.~9, pp. 3936--3964, 2006.

\bibitem{Jin_ACMISM05_cacheWCDN}
S.~Jin and L.Wang, ``Content and service replication strategies in multihop
  wireless mesh networks,'' in \emph{in Proc. 8th ACM Int. Symp. Model., Anal.
  Simul. Wireless Mobile Syst., Montreal, QC, Canada}, Oct. 2005, pp. 79--86.

\bibitem{Bondy_Elsevier76_GraphTheory}
J.~A. Bondy and U.~Murthy, \emph{Graph Theory with Applications}.\hskip 1em
  plus 0.5em minus 0.4em\relax New York: Elsevier, 1976.

\bibitem{Liu_10sTSP_Fairness_rate_polit_WF}
A.~Liu, Y.~Liu, H.~Xiang, and W.~Luo, ``{MIMO B-MAC} interference network
  optimization under rate constraints by polite water-filling and duality,''
  \emph{IEEE Trans. Signal Processing}, vol.~59, no.~1, pp. 263 --276, Jan.
  2011.

\end{thebibliography}
\end{document}